\documentclass[fleqn,usenatbib]{mnras}

\usepackage{newtxtext,newtxmath}

\usepackage[T1]{fontenc}

\DeclareRobustCommand{\VAN}[3]{#2}
\let\VANthebibliography\thebibliography
\def\thebibliography{\DeclareRobustCommand{\VAN}[3]{##3}\VANthebibliography}

\usepackage{graphicx}	
\usepackage{amsmath}	
\usepackage[table,xcdraw]{xcolor}
\usepackage{gensymb}
\usepackage{placeins}
\usepackage{longtable}
\usepackage{caption}
\usepackage{multirow}
\usepackage{graphicx}
\definecolor{lightgray}{gray}{0.9}
\usepackage{siunitx}
\usepackage{hhline}

\title[Origin of gas absorption in debris discs]{A search for circumstellar gas in pre-main-sequence debris discs using absorption spectroscopy}

\author[K. M. Szewczyk et al.]{
Karolina M. Szewczyk$^{1}$\thanks{E-mail: py19k2ms@leeds.ac.uk},
D. P. Iglesias$^{1,2}$,
and O. Panić$^{1}$
\\
$^{1}$ School of Physics \& Astronomy, University of Leeds, Sir William Henry Bragg Building, Woodhouse Lane, Leeds LS2 9JT, UK\\
$^{2}$ Astrophysics Research Cluster, School of Mathematical and Physical Sciences, The University of Sheffield, Hounsfield Road, Sheffield, S3 7RH, UK}

\date{Accepted XXX. Received YYY; in original form ZZZ}

\pubyear{2025}

\begin{document}
\label{firstpage}
\pagerange{\pageref{firstpage}--\pageref{lastpage}}
\maketitle

\newcommand{\DI}{\color{red} DI: }
\newcommand{\new}{\color{teal}}

\begin{abstract}
Gas in debris discs is thought to be either inherited from the protoplanetary stage or released from the solid, rocky content of planetesimal belts. Its presence can impact planetary atmospheres and their potential for habitability, which stresses the need to ascertain its origin and composition. Most detections to date are around main-sequence stars, with only a few gas-bearing debris discs identified around pre-main-sequence stars, mainly through millimetre CO line searches. We investigate narrow gas absorption features superimposed on the photospheric Ca\,\textsc{ii} K \& H and Na\,\textsc{i} D1 \& D2 lines in a sample of 125 pre-main sequence and 5 relatively young (<\,17\,Myr) stars. All stars are associated with IR excess emission indicative of presence of a debris disc. By comparing their residual spectra (photosphere-subtracted) to those of nearby stars, interstellar cloud velocities, and stellar radial velocities, we test whether interstellar absorption is the culprit and ascertain circumstellar gas origin. Using these methods, out of the 130 targets, we identified two new gas-bearing debris discs: TYC\,7879-1373-1, which exhibits stable absorption, and HIP\,30414, which shows variable gas absorption features linked likely to ongoing accretion. Both these systems are pre-main-sequence stars younger than 5\,Myr. TYC\,6822-283-1 has absorption features of inconclusive origin. This study increases the number of currently known very young (<\,10\,Myr) debris discs with circumstellar gas to eight, paving the path to future systematic studies of objects caught in transition from protoplanetary to debris disc stages. 
\end{abstract}

\begin{keywords}
circumstellar matter -- stars: pre-main-sequence -- protoplanetary discs -- planetary systems
\end{keywords}



\section{Introduction}

\begin{figure}
    \centering
    \includegraphics[width=0.95\columnwidth]{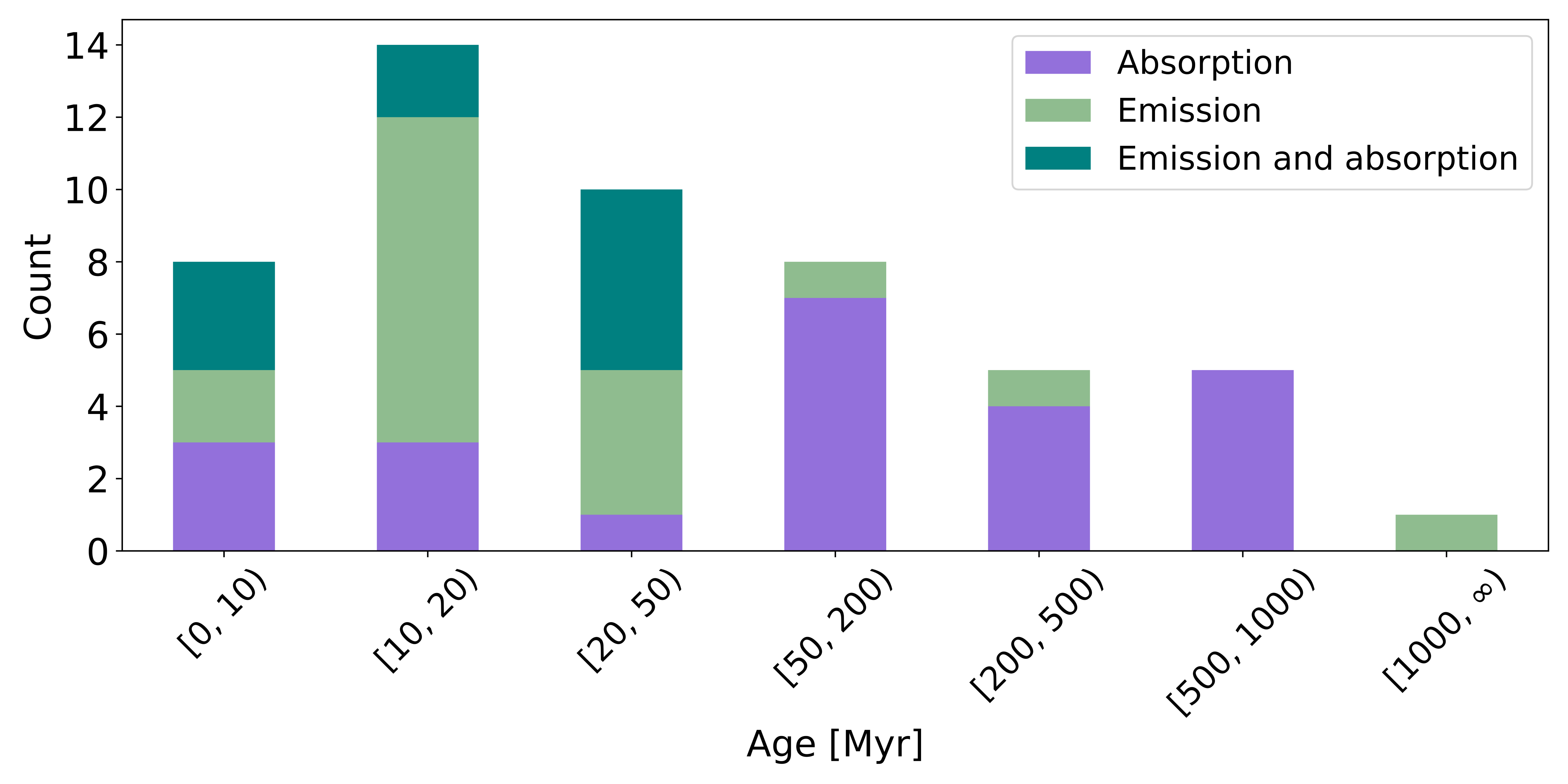}
    \caption{Histogram of all gas detections in debris discs through either emission or absorption, divided into bins of age, including HIP\,30414 and TYC\,7879-1373-1. Objects with gas emission and their ages were adopted from \citet{Szewczyk2025} (Table A1), based on the data from \citet{Lovell2021ALMADispersal}, \citet{Miley2018Unlocking141569}, \citet{Rebollido2022The36546}, \citet{Matra2019On7}, \citet{Lieman-Sifry2016DebrisALMA}, \citet{Schneiderman2021}, \citet{Hales2022}, \citet{Moor2017MolecularStars}, \citet{Marino2016ExocometaryRing}, \citet{Dent2014MolecularDisk}, \citet{Riviere-Marichalar2014GasObservatory}, \citet{Moor2019New32297}, \citet{Kospal2013ALMA21997}, \citet{Pericaud2017TheDisks}, \citet{Matra2017DetectionComets}, and \citet{Marino2017}. We additionally included HD\,9985, HD\,145101, HD\,152989, HD\,155853, and HD\,170116 from \citet{Moor2025}. Objects with gas absorption and their ages used in this figure were adopted from \citet{Iglesias2020SearchingDisks} (Table\,1.1), based on data from \citet{Chen2003TheHerculis}, \citet{Lagrange-Henri1990SearchStar.}, \citet{Abt1973RotationDwarfs.}, \citet{Welsh2013CircumstellarExocomets}, \citet{Koubsky1993ComingHerculis.}, \citet{Iglesias2018}, \citet{Iglesias2019AnDisc}, \citet{Welsh1998Beta85905}, \citet{Welsh2015TheAbsorption}, \citet{Montgomery2012DetectionStars}, \citet{Welsh2018FurtherDiscs}.}
    \label{fig:histogramAllDetections}
\end{figure}

The detection of gas in debris discs remains relatively rare, although recent discoveries suggest it may not be as uncommon as once thought. Debris discs are circumstellar discs of dust and planetesimals typically found around main-sequence (MS) stars \citep[although not limited to them,][]{Veras2020ConstrainingStars}. Protoplanetary discs, which are the earlier evolutionary stage, are gas-rich objects 
formed from the initial molecular cloud collapse, whose mass is dominated by gas, with gas-to-dust mass ratio of $\sim$100 \citep{PPVII2023}. During the transition from protoplanetary to debris phase, most of the gas (and dust) is removed through a combination of photoevaporation, accretion onto the star, and planet formation, typically within $\sim$3-11\,Myr but depending on the stellar mass of the parent star \citep{Ribas2015,Hernandez2007,Haisch2001}. The primordial disc becomes depleted, leaving behind larger bodies, which will continue to undergo collisions, generating dust that sustains the observed debris disc \citep{Wyatt2008EvolutionDisks}. Unlike protoplanetary discs, debris discs generally do not contain much gas and used to be considered devoid of it \citep[e.g., in][]{Backman1993}. However, the discovery of gas in the debris disc of $\beta$ Pic \citep{Hobbs1985ThePictoris} challenged this belief and initiated the efforts to understand the physical mechanisms that allow the gas to be present in the later stages of planetary system evolution. To date, the debris disc around $\beta$\,Pic contains the widest range of detected gas species, including Ca, Na, CO, C, O, Fe, and many more \citep[for a complete list see][]{Iglesias2025}. This transient gas has since been shown to be primarily produced by exocomets - minor bodies that sublimate as they approach the star \citep{Dent2014MolecularDisk, Kral2016APictoris}.

Gas has now been detected in nearly 50 debris discs, with ages ranging from $\sim$2 to $\sim$1500\,Myr (1-3 Myr NO Lup, \citealt{Lovell2021ALMADispersal}; 1-2 Gyr $\eta$ Crv, \citealt{Marino2017}) through either emission \citep[e.g.,][]{Lieman-Sifry2016DebrisALMA, Moor2017MolecularStars,Hales2022,Cataldi2023PrimordialALMA} or absorption \citep[e.g.,][]{Lagrange-Henri1990SearchStar.,Montgomery2012DetectionStars,Iglesias2018,Iglesias2019AnDisc}, summarised in Figure\,\ref{fig:histogramAllDetections}. These systems span a wide range of CO abundances. Some, like $\beta$\,Pic, only contain low levels of CO, while others show CO gas masses comparable to those in protoplanetary discs. While emission studies allow for direct observations of gas, identifying it through absorption requires less telescope time \citep{Iglesias2018}, hence is a time-effective method. The gas species commonly detected differ between the two methods. CO and C\,\textsc{i} are primarily observed in emission, and CO is currently the only molecule detected in the millimetre regime \citep[e.g.,][]{Kral2020Survey129590, Hales2022, Lovell2021ALMADispersal, Lieman-Sifry2016DebrisALMA}; Ca\,\textsc{ii} K \& H and Na\,\textsc{i} D1 \& D2 doublets are the most commonly used in absorption spectroscopy \citep[e.g.,][]{Iglesias2018, Hales2017, Welsh2015TheAbsorption}, although CO and C\,\textsc{i} have also been detected in absorption in a few systems \citep[e.g.,][]{Roberge2000, Brennan2024}. The wavelengths probed by these two methods also differ. Emission studies typically target sub-millimetre and millimetre transitions of cold molecular gas using radio interferometers, while absorption studies detect ultraviolet and optical absorption lines seen against the stellar continuum. Nonetheless, both methods are crucial for our understanding of the evolution of debris discs and the processes that shape planetary systems.

Over the past decade, numerous studies have searched for gas in debris discs by analysing Ca\,\textsc{ii} absorption lines. \citet{Hales2017} examined spectra of 16 debris disc host stars and found four stars showing narrow Ca\,\textsc{ii} or Na\,\textsc{i}\,D absorption features. However, a comparison with stellar radial velocities, interstellar clouds, column densities, and neighbouring stars suggested that only one system, HD\,110058, hosted circumstellar gas. Similarly, in a survey of 301 debris discs, 23 of which exhibited narrow absorption features, \citet{Iglesias2018} identified three systems with circumstellar gas: HD\,110058, with stable absorption, as well as HR\,4796 and c\,Aql, with variable absorption features. 

Two main gas origin scenarios have been proposed for debris discs: primordial and secondary. In case of CO-poor discs, like $\beta$\,Pic, the origin of the gas is well established as secondary - produced within the debris disc, in a similar manner to dust, through, e.g., collisions or sublimation of planetesimals \citep{Kral2017,Kral2019ImagingDiscs,Hughes2018}. In these systems the gas has to be of secondary origin, as the low CO levels must be continuously replenished, since the CO is rapidly photodissociated by the interstellar radiation field on a timescale of $\sim$130\,yr when unshielded \citep{Heays2017,Marino2022VerticalLifetime}, so any primordial reservoir would long have been destroyed. Less is known in the case of CO-rich debris discs, in which both origin scenarios have been proposed. In the primordial scenario, the CO is a remnant of the protoplanetary disc, surviving from the initial collapse of the molecular cloud \citep{Wyatt2015,Nakatani2021PhotoevaporationRemnants,Smirnov-Pinchukov2022LackGas}. Most gas-bearing debris discs have been found around young ($\lesssim$40\,Myr), A-type stars \citep[][see also Figure\,\ref{fig:histogramAllDetections}]{Hughes2018}, supporting the primordial gas origin as these stars may be more likely to retain circumstellar gas from earlier evolutionary stages \citep{Nakatani2021PhotoevaporationRemnants}. In pre-MS A-type stars, X-ray radiation is weaker than in low mass stars, while far-ultraviolet (FUV) radiation is the dominant photoevaporation mechanism. However, grain growth may inhibit FUV photoevaporation, potentially allowing gas to persist into the debris stage \citep{Nakatani2021PhotoevaporationRemnants}. In the absence of small dust grains, extreme-ultraviolet (EUV) radiation also becomes inefficient, allowing some of the gas to persist into the debris phase, where it may influence system dynamics \citep{Nakatani2023}. Detections around lower mass stars are indeed rarer \citep{Wyatt2015}, perhaps due to their stronger X-ray and FUV radiation driving more efficient photoevaporation. While CO in older debris discs, such as the 440\,Myr Fomalhaut, is most likely secondary \citep{Matra2017DetectionComets}, the youngest systems may still retain primordial gas. 

Very few gas-bearing debris discs have been found around pre-MS stars, particularly those younger than $\sim$10\,Myr. Systems such as NO Lup \citep[1-3\,Myr,][]{Lovell2021RapidDisc}, HD\,44892 \citep[2.1\,Myr,][]{Szewczyk2025}, and HD\,141569 \citep[5\,Myr,][]{Miley2018Unlocking141569} remain rare, although the gas in NO\,Lup is suspected to be wind-driven rather than from a disc. HD\,44892 is a pre-MS A-type star, also known as HIP\,30414, and a part of the sample we studied  - we report the first evidence of gas absorption towards this disc in later sections. Targeting such young candidates is crucial for probing the poorly understood transitional stage from protoplanetary to debris discs, as it allows us to trace how the gas evolves in the first few million years after disc dispersal and how it may influence early planet formation and system architecture.

The origin of gas in the CO-rich debris discs is still not fully understood. While their dust masses are more typical of debris discs, their CO gas masses can be of similar magnitude to those of protoplanetary discs. \citet{Pericaud2017TheDisks} first introduced the term 'hybrid discs' for these systems, defining them as having gas of primordial origin but secondary dust. However, since then, some studies have shown that secondary processes, such as CO release from colliding or sublimating planetesimals, may also produce the high CO gas masses observed \citep[e.g.,][]{Kral2017, Kral2019ImagingDiscs}, challenging the primordial gas interpretation. As the CO gas origin in these systems remains uncertain, we adopt the term 'hybrid disc candidates' only for the candidates identified by \citet{Iglesias2023}, in line with their classification and terminology. The youngest of CO-bearing debris discs are key for understanding how long gas can persist and influence planet formation - while giant planets are expected to form during the protoplanetary phase, terrestrial planet formation may continue in the debris stage and be shaped by the presence of gas \citep{Kral2020FormationAccretion}. HD\,141569 is a well-studied example of a CO-rich disc \citep['hybrid disc' in][]{Iglesias2023}, with a CO gas mass comparable to protoplanetary discs but a dust mass more typical of debris discs. Based on similarities in fractional infrared (IR) excess to HD\,141569, \citet{Iglesias2023} identified 17 very young objects as 'hybrid disc candidates'. Our study builds upon their work by searching for new detections of circumstellar gas in both their 'hybrid candidates' and debris disc systems (all <\,17\,Myr), for a total sample of 130 targets. 

\section{Sample selection, observations and data reduction}

\subsection{Sample selection}
\label{sec:Sample}

Studying pre-MS stars is key to understanding the transition from protoplanetary to debris discs. \citet{Iglesias2023} conducted a large spectroscopic survey of intermediate-mass (1.5\,-\,3.5\,M$_{\odot}$) pre-MS stars using X-Shooter on the Very Large Telescope (VLT), aiming to identify new young stars exhibiting disc excess. From their full sample, they identified 135 objects showing infrared (IR) excess at 12\,$\mu$m, indicative of warm dust in a circumstellar disc. These systems were classified using their fractional 12\,$\mu$m excess, with protoplanetary discs defined as those with $\frac{F_{12\mu m}}{F_{*}}>25$ and debris discs with $\frac{F_{12\mu m}}{F_{*}}<2$ \citep{Wyatt2015, Iglesias2023}.

From the 135 sources identified by \citet{Iglesias2023}, we exclude the five systems whose excesses fall in the protoplanetary disc regime, leaving a working sample of 130 objects. Of these, 22 had suitable archival high-resolution spectra, while for the remaining 108 we obtained new Ultraviolet and Visual Echelle Spectrograph (UVES; \citealt{Dekker2000}) observations. The sample spans stellar masses of 1.51\,-\,3.46\,M$_\odot$, distances of 54.8\,-\,292.0\,pc, and ages between 1.61 and 16.94\,Myr. Based on their 12\,$\mu$m fractional excesses, 113 of the 130 stars fall within the debris disc regime, while 17 are found in the region between debris and protoplanetary discs, classified as 'hybrid disc candidates' in \citet{Iglesias2023} based on \citet{Wyatt2015}. The sample consists of B-, A-, and F-type stars, which have fewer and broader spectral lines \citep{Reffert2015PreciseStars}, making narrow absorption features easier to identify. A-type stars in particular have been found to host gas-bearing debris discs more frequently than other spectral types, with $\sim$70\% of known detections occurring around them \citep{Moor2017MolecularStars}. Moreover, given their young ages ($<$17\,Myr), all 130 systems, whether already debris or caught in the transition from protoplanetary to debris disc stages, are plausible hosts of circumstellar gas, which may be primordial, secondary, or mixed, as most detections are around younger debris disc systems (see Figure\,\ref{fig:histogramAllDetections}).

\subsection{Observations}\label{sec:observations}
 
We obtained the high-resolution spectra for the 108 stars in our sample which did not have suitable archival data available using UVES at the VLT in Cerro Paranal, Chile. Observations were carried out between May 2022 and February 2024 (ESO periods 109, 110 and 112)\footnote{ESO programmes 109.23K8.001, 110.23YZ.001, and 112.262P.001.  PI: Iglesias.}. Spectra were obtained using the 0.6'' slit and the dichroic DIC1 setup centered at 390\,nm and 580\,nm, respectively, for the BLUE and RED arms. The remaining 22 stars already had high-resolution (R\,$>$\,40000) spectroscopic observations in the ESO archive\footnote{\url{http://archive.eso.org}} from spectrographs such as UVES, HARPS \citep{Mayor2003SettingHARPS}, and FEROS \citep{Kaufer1999CommissioningLa-Silla.}, at the La Silla Observatory. Therefore, for these stars we used their available archival spectra instead of obtaining new observations. Observations of our sample are summarized in Table~\ref{tab:ObservationDetails}.

In addition, for each star in the sample, we identified comparison stars based on their nearby location on the sky (see Section\,\ref{sec:MethodsCharacterisation}) and for these we searched the ESO archive for high resolution spectroscopic data. We retrieved optical spectra from UVES, FEROS or HARPS spectrographs of stars within a maximum of 5$\degree$ of our targets, resulting in a sample of 134 comparison stars. Wherever possible, we prioritised stars with similar distances from the Sun, smallest angular separation from target, and with B, A, or F spectral types, since narrow absorption features are more easily identified in these stellar types. Where we did not find suitable comparison stars in the ESO archive, we conducted additional observations\footnote{ESO programme 113.26V6.001. PI: Iglesias.} with UVES of 38 stars using the same observing setup, to ensure that each target had at least one well-suited comparison star, ideally within 1$\degree$ separation. 
These additional observations are listed in Table~\ref{tab:ObservationDetailsNearby}.

\subsection{Data reduction}

All spectra of both science and comparison stars have been reduced with the data reduction software corresponding to each instrument. We reduced our UVES observations using the \texttt{EsoReflex} pipeline \citep{Freudling2013} version 2.11.5. We applied a barycentric radial velocity correction to all UVES spectra a posteriori, as this is not included in the reduction pipeline. This is important for the determination of the radial velocity of the features and, thus, their diagnostic. For the RED arm spectra, in the case of UVES, and where pertinent, for the HARPS and FEROS data, we used Molecfit \citep{Smette2015Molecfit:Correction,Kausch2015Molecfit:Correction} to correct for tellurics. The red wavelengths are highly contaminated from telluric lines due to mainly water vapour, O$_2$, and O$_3$ in the Earth's atmosphere. In particular, the Na\,\textsc{i} doublet is significantly affected by telluric lines contamination and, therefore, it is crucial to eliminate these lines when performing the analysis of the sodium lines.

\section{Methods}\label{MethodsNarrowFeatures}

\subsection{Identifying narrow absorption features} 

Gas detected through absorption spectroscopy does not necessarily originate from a circumstellar disc. Such features can arise from any gas in the line of sight between the observer and the star \citep{Iglesias2018}, including both the interstellar medium (ISM) clouds and circumstellar material. Determining the origin of these features is therefore crucial. We searched for narrow absorption features in the Ca\,\textsc{ii} K \& H (at 3933.663\,{\AA} \& 3968.469\,{\AA}) and Na\,\textsc{i} D1 \& D2 (at 5895.924\,{\AA} \& 5889.950\,{\AA}) lines, which are superimposed on the star's broad photospheric absorption profiles. These additional components are far narrower than the stellar lines in early-type stars, typically only a few km\,s$^{-1}$ wide, hence cannot be photospheric in nature and must instead result from the absorption of stellar light by either interstellar or circumstellar gas \citep{Hales2017}. For later-type stars, the photospheric lines are intrinsically narrower, making it more difficult to distinguish them from non-photospheric components. The Ca\,\textsc{ii} K line in particular is a well-known tracer of circumstellar (and interstellar) gas and is widely used to identify gas-bearing debris discs. This line is most sensitive to the transit of gaseous clouds, such as those produced by exocomets \citep{Kiefer2014}. In contrast, the Ca\,\textsc{ii} H line absorption is usually half as strong as the Ca\,\textsc{ii} K line \citep{Iglesias2018}, hence often undetectable in low signal-to-noise spectra \citep{Bendahan-West2025}. The Na\,\textsc{i} doublet is efficiently photoionised in the stellar UV radiation field, which is stronger than that of the ISM. Thus, there is often very little neutral sodium present in the circumstellar disc compared to ionised calcium (Ca\,\textsc{ii}) \citep{Redfield2002}. While the Na\,\textsc{i} doublet is therefore a useful tracer of ISM gas, where weaker radiation field allows it to remain in its neutral state, it is expected to be much less abundant inside a debris disc \citep[but can still be detectable, e.g., like in $\beta$\,Pic,][]{Olofsson2001}.

Many of our targets had multiple observational epochs available, either from archival data or our UVES observations (see Table\,\ref{tab:ObservationDetails} for details). To ensure consistent continuum levels across the different epochs, we normalised each spectrum by rescaling the flux data within a radial velocity window of $\pm$100\,km\,s$^{-1}$ centred on a specific spectral line, such that the maximum flux value was set to 1. Where multiple epochs were available, we combined the normalised spectra by computing the median flux at each wavelength point, improving the signal-to-noise ratio and isolating features that were consistently present across most observed epochs. 

For most targets, we modelled the photospheric absorption using a polynomial (spline) fit, with the order of the polynomial selected interactively for each target to best reproduce the shape of the photospheric absorption profile. This method provides a good approximation of the stellar spectrum, particularly in the cases where the narrow absorption features superimposed on top of it are stable across the observations. However, one object in our sample (discussed in Section\,\ref{sec:HIP30414var}), showed significant variability in its narrow absorption features. This made it difficult to accurately identify the photospheric absorption, making the spline fit unreliable for that object. To account for this, we fitted this single target's spectra using a Kurucz model \citep{CastelliKurucz2003}, which is a synthetic stellar atmosphere model. While this model is generally more accurate, it is also significantly more complex and time consuming to produce, making it impractical for most targets unless necessary. Since other objects in our sample do not exhibit significant variability or complexity, a spline fit was sufficient to model their stellar photospheres. The photosphere fits for the individual systems in which circumstellar gas is discussed in detail can be found in Appendix \ref{Appendix:B}.

In both cases (spline and Kurucz model fit), the spectra were then divided by this fit to obtain continuum-subtracted spectra, isolating the narrow absorption features. For absorption lines showing a radial velocity shift in a cyclic manner over multiple epochs, we investigated the possibility of that object being a binary system. For this, we inspected the spectra for signs of line-doubling or cyclic variability, and searched the literature and relevant catalogues, e.g., SIMBAD \citep{Simbad}, VizieR \citep{vizier}, for existing binarity classifications.

\subsection{Characterising the origin of narrow absorption features}\label{sec:MethodsCharacterisation}

\begin{table*}
\renewcommand{\arraystretch}{1.5}
\caption{\label{tab:DiscussedObjectsParams}Parameters of objects discussed in this paper as having potential circumstellar gas, based on \citet{Iglesias2023}. Full list of objects from the \citet{Iglesias2023} sample used in this study can be found in Table \ref{tab:AllObjects}. TYC\,6822-283-1 is marked with $^{i}$ as the origin of detected gas is inconclusive.}
\begin{tabular}{|l|l|l|l|l|l|l|l|}
\hline
Name & Mass [$M_{\odot}$] & Luminosity [$L_{\odot}$] & SpT & Temperature [K] & Distance [pc] & Age [Myr] & $F/F_{star}$ at 12\,$\mu$m\\ \hline
HIP\,30414 & 2.87$^{+0.67}_{-0.41}$ & 60.0~$\pm$~20.0	& A9/F0IV & 7500~$\pm$~500 & 190.39 & 2.10$^{+1.20}_{-1.00}$ & 7.86$^{+0.11}_{-2.27}$ \\ \hline
TYC\,7879-1373-1 & 2.66$^{+0.10}_{-0.23}$ & 60.0~$\pm$~10.0 & A0V & 10500~$\pm$~500 & 225.64 & 3.30$^{+1.00}_{-0.30}$ & 1.54$^{+0.13}_{-0.11}$ \\ \hline
TYC\,6822-283-1$^{i}$ & 2.50$^{+0.18}_{-0.24}$ & 50.0~$\pm$~10.0 & A0V & 10700~$\pm$~700 & 187.69 & 4.00$^{+	2.00}_{-1.00}$ & 6.21$^{+0.55}_{-0.52}$ \\ \hline
\end{tabular}
\end{table*}

We characterised the narrow absorption features detected in the Ca\,\textsc{ii} K\& H and Na\,\textsc{i} D1 \& D2 lines by fitting them with Gaussian profiles, using either single or multiple components to model the continuum-subtracted spectra. The root-mean-square (RMS, $\sigma$) of the continuum was used to estimate the noise level, and only features with depths exceeding $3\sigma$ were considered detections. For each detected feature, we measured the radial velocity and intensity (depth) from its Gaussian fit. After characterising the narrow features, our aim was to determine whether they were of interstellar or circumstellar origin. To do this, we followed the methodology from \citet{Iglesias2018}, which included multiple diagnostic methods: i) comparison with known interstellar clouds along the line of sight; ii) comparison with narrow absorption features in spectra of nearby stars; iii) comparison with the stellar radial velocity, since stable gas in Keplerian rotation should approximately match this velocity; and iv) searching for variability across different epochs, which can indicate transient circumstellar gas.

To assess whether any of the detected narrow absorption features could be attributed to interstellar clouds, we used the \citet{Redfield2008} database\footnote{\url{http://lism.wesleyan.edu/LISMdynamics.html}}, which catalogues 15 local interstellar clouds within 15\,pc of the Sun and provides their predicted radial velocities. For each target, we used it to identify clouds lying directly along the line of sight or passing within 20$\degree$ of it. For example, we found that TYC\,6822-283-1 has a sight line transverse to the G cloud (-28.39\,$\pm$\,1.13\,km\,s$^{-1}$) and passes within 20$\degree$ of the Gem and Aql clouds (-29.78\,$\pm$\,1.07\,km\,s$^{-1}$ and -42.27\,$\pm$\,1.17\,km\,s$^{-1}$, respectively). However, because this database only covers the local ISM, it is incomplete for our most distant sources (up to $\sim$300\,pc). We therefore also searched for interstellar contamination in the spectra of comparison stars located in similar sky directions. 

The spectra of comparison stars were obtained from the ESO archive and our own observations (see Section\,\ref{sec:observations} for details). We processed their spectra in the same manner as in our target sample, and calculated the angular separations between objects using TopCat \citep{TopCat}. For each of our targets, we plotted only the residual flux as a function of radial velocity and directly compared it with the residual spectra of the nearby comparison stars. If a narrow absorption feature appeared at the same radial velocity (within uncertainties) in multiple neighbouring stars, we classified it as likely interstellar in origin.

To further investigate the nature of each detected gas absorption feature, we compared their measured radial velocity with the stellar radial velocities reported by \citet{Iglesias2023}, derived from X-shooter data. Gas rotating in a circumstellar disc is expected to match that velocity, and significant deviations can suggest an interstellar origin instead. For the three objects showing possible evidence of circumstellar gas, we calculated more precise stellar radial velocities using our higher-resolution UVES spectra to improve the accuracy of this comparison. Infall- and outflow-driven absorption, commonly observed in earlier evolutionary stages \citep[Class 0/1, ][]{Lada1987}, is not expected in our targets, which are more evolved and beyond that phase. Any gas detected in our targets will likely be rotating around the star and therefore have a projected radial velocity close to that of the star. However, transient phenomena such as accretion, disc winds, or exocomets may produce absorption offset from the stellar velocity, which we take into consideration when analysing individual systems.

Following the approach of \citet{Iglesias2018,Iglesias2023winds}, we calculated the apparent column densities of narrow absorption features for the objects of interest using the methodology described by \citet{Somerville1988}. Oscillator strength values ($f$) were taken from \citet{Morton1991}, and equivalent width values were derived from our Gaussian profile fits. We also calculated the apparent column density ratios (Ca\,\textsc{ii}\,/\,Na\,\textsc{i}, where the Ca\,\textsc{ii} and Na\,\textsc{i} column densities correspond to the sums of K \& H and D1 \& D2 lines, respectively, see Table\,\ref{tab:ColumnDensity}), which can serve as a diagnostic tool for assessing the origin of the absorption features. Ratios $\lesssim$\,1 are typically associated with ISM gas, suggesting interstellar origin - although they do not fully exclude circumstellar nature \citep{Iglesias2018}.

\section{Results}

\begin{table*}
\renewcommand{\arraystretch}{1.5}
\caption{\label{tab:ColumnDensity}Parameters of absorption features found in spectra of each object, stated for each line, including the feature's radial velocity, apparent column density, and equivalent width. Last column is the apparent column ratio of Ca\,\textsc{ii} to Na\,\textsc{i}. Parameters of features of suspected circumstellar origin are shaded. TYC\,6822-283-1 is marked with $^{i}$ as the origin of detected gas is inconclusive. $^{UL}$ denotes an upper limit where no feature is detected. Parameters of HIP\,30414 are presented for both of its observed epochs to show variability between them. $^{*}$Uncertainties are of order 2-4\% for Ca\,\textsc{ii} and 1-2\% for Na\,\textsc{i}.} 
\resizebox{\textwidth}{!}{%
\begin{tabular}{|c|c|c|c|c|c|c|c|c|c|c|c|c|c|}
\hline
& \multicolumn{3}{c|}{Ca\,\textsc{ii}\,K$^{*}$} & \multicolumn{3}{c|}{Ca\,\textsc{ii}\,H$^{*}$} & \multicolumn{3}{c|}{Na\,\textsc{i}\,D1$^{*}$} & \multicolumn{3}{c|}{Na\,\textsc{i}\,D2$^{*}$} & \\ \hline
Name & RadV & log$_{10}$N & EW & RadV & log$_{10}$N & EW & RadV & log$_{10}$N & EW & RadV & log$_{10}$N & EW & $\frac{\text{N[CaII]}}{\text{N[NaI]}}$ \\ 
& \scriptsize{[km\,s$^{-1}$]} & \scriptsize{[cm$^{-2}$]} & \scriptsize{[m$\si{\angstrom}$]} & \scriptsize{[km\,s$^{-1}$]} & \scriptsize{[cm$^{-2}$]} & \scriptsize{[m$\si{\angstrom}$]} & \scriptsize{[km\,s$^{-1}$]} & \scriptsize{[cm$^{-2}$]} & \scriptsize{[m$\si{\angstrom}$]} & \scriptsize{[km\,s$^{-1}$]} & \scriptsize{[cm$^{-2}$]} & \scriptsize{[m$\si{\angstrom}$]} & \\ \hline
\rowcolor{lightgray}
\cellcolor{white} HIP\,30414 \scriptsize{epoch 1} & \scriptsize{\textit{variable}} & 12.43 & 232.38 &  \scriptsize{\textit{variable}} & 12.53 & 148.26 & \scriptsize{\textit{variable}} & 12.14 & 133.73 & \scriptsize{\textit{variable}} & 11.89 & 150.10 & 2.85 \\
 & 22.47 & 11.73 & 47.11 & 21.56 & 11.80 & 27.40 & 22.66 & 12.14 & 132.75 & 22.62 & 11.88 & 146.57 & 0.55 \\
\hhline{~-------------}
\rowcolor{lightgray}
\cellcolor{white} HIP\,30414 \scriptsize{epoch 2} & \scriptsize{\textit{variable}} &  12.42 & 229.62 & \scriptsize{\textit{variable}} & 12.50 & 138.86 & \scriptsize{\textit{variable}} & 12.14 & 133.54 & \scriptsize{\textit{variable}} & 11.88 & 147.94 & 2.74 \\
& 23.09 & 11.75 & 48.68 & 22.53 & 11.83 & 29.92 & 22.69 & 12.14 & 132.36 & 22.68 & 11.88 & 145.58 & 0.59 \\ \hline
\rowcolor{lightgray}
\cellcolor{white} TYC\,7879-1373-1 & -11.46 & 10.44 & 2.36 & -13.30 & 10.71 & 2.25 & - & 10.21$^{UL}$ & 1.57$^{UL}$ & - &  9.71$^{UL}$ & 0.98$^{UL}$ & 3.67 \\ 
& -3.94 & 11.46 & 25.06 & -4.15 & 11.55 & 15.49 & -4.41 & 11.97 & 90.66 & -4.59 & 11.68 & 92.54 & 0.45 \\ 
& 1.43 & 11.42 & 12.86 & 1.74 & 11.58 & 7.41 & 1.25 & 12.03 & 103.64 & 1.06 & 11.78 & 117.68 & 0.38 \\ \hline
TYC\,6822-283-1$^{i}$ & -16.01 & 11.27 & 16.14 & -16.92 & 11.50 & 13.77 & -16.63 & 10.56 & 3.54 & -15.82 & 10.96 & 17.49 & 3.98 \\
& - & 10.46$^{UL}$ & 2.49$^{UL}$ & - & 10.80$^{UL}$ & 2.78$^{UL}$ & -11.66 & 11.06 & 11.21 & -11.33 & 11.15 & 27.11 & 0.35 \\
& -5.53 & 10.82 & 5.77 & -5.36 & 11.14 & 6.09 & -5.47 & 11.87 & 71.39 & -5.39 & 11.65 & 87.43 & 0.17 \\
& 0.31 & 11.61 & 35.30 & 0.22 & 11.71 & 22.64 & 1.02 & 11.33 & 20.85 & 1.20 & 11.33 & 41.45 & 2.14 \\
\rowcolor{lightgray}
\cellcolor{white} & 7.16 & 11.45 & 24.54 & 6.74 & 11.57 & 16.15 & 7.42 & 10.92 & 8.06 & 7.55 & 10.88 & 14.54 & 4.07 \\
\rowcolor{lightgray}
\cellcolor{white} & 12.29 & 10.29 & 1.71 & 10.97 & 10.82 & 2.91 & - & 10.21$^{UL}$ & 1.57$^{UL}$ & - & 9.31$^{UL}$ & 0.39$^{UL}$ & 4.68 \\ \hline
\end{tabular}
}
\end{table*}

\begin{figure*} 
\centering
\includegraphics[width=1.8\columnwidth]{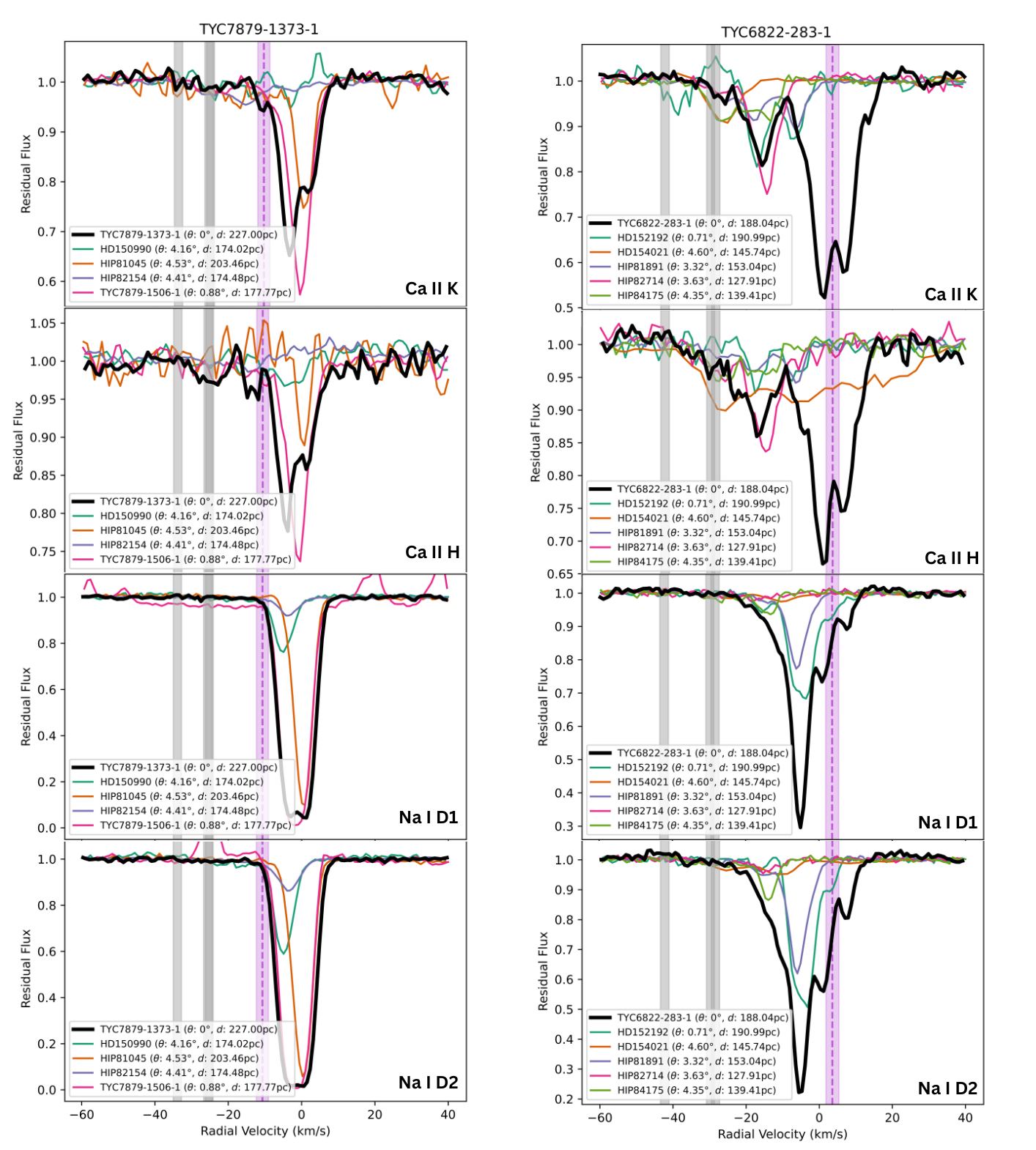}
\caption{\label{fig:TYC7879-1373-1_TYC6822-283-1}\textbf{Left:} Residual flux vs radial velocity [km\,s$^{-1}$] showing three isolated narrow absorption features in the Ca\,\textsc{ii} K \& H and Na\,\textsc{i} D1 \& D2 lines of TYC\,7879-1373-1 (in black), which, in Ca\,\textsc{ii} K, are -11.33\,km\,s$^{-1}$ (7.4$\sigma$), -4.02\,km\,s$^{-1}$ (42.6$\sigma$), and 2.04\,km\,s$^{-1}$ (27.9$\sigma$), respectively. The known velocities of clouds in the line of sight are plotted in grey. Comparison stars with their narrow absorption features are plotted in various colours, with their distances and angular separations from TYC\,7879-1373-1 stated. The pink region corresponds to the radial velocity of TYC\,7879-1373-1, -10.68\,$\pm$\,1.64\,km\,s$^{-1}$. \textbf{Right: }Same but for TYC\,6822-283-1. Radial velocity of TYC\,6822-283-1 is 3.60\,$\pm$\,1.77\,km\,s$^{-1}$. The five narrow absorption features detected in Ca\,\textsc{ii} K line are -16.01\,km\,s$^{-1}$ (12.9$\sigma$), -5.53\,km\,s$^{-1}$ (10.2$\sigma$), 0.31\,km\,s$^{-1}$ (37.0$\sigma$), 7.16\,km\,s$^{-1}$ (36.5$\sigma$), and 12.29\,km\,s$^{-1}$ (10.7$\sigma$).}
\end{figure*}

\begin{figure*} 
\centering
\includegraphics[width=1.8\columnwidth]{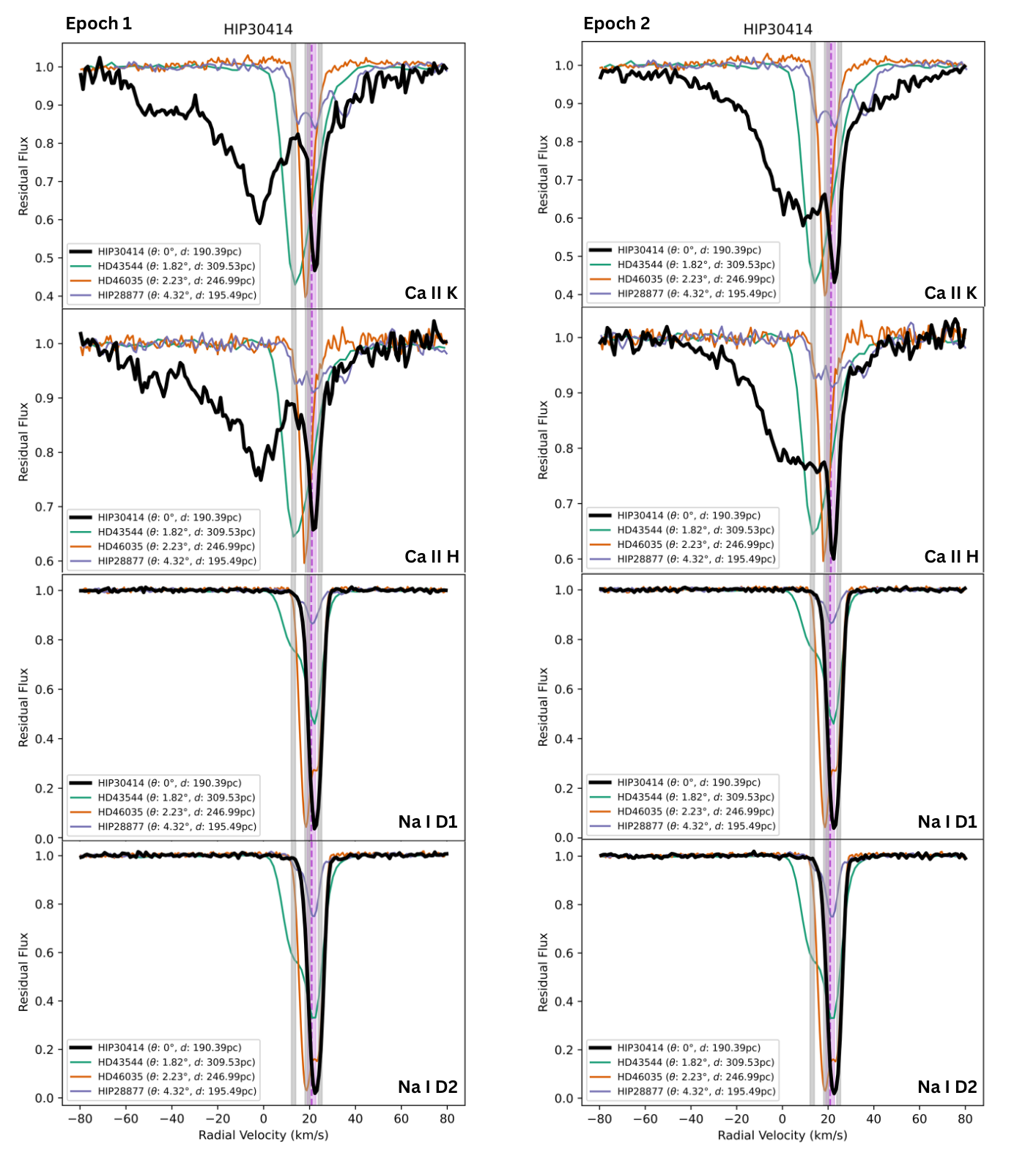}
\caption{\label{fig:HIP30414}\textbf{Left:} Residual flux vs radial velocity [km\,s$^{-1}$] for epoch 1, showing variable isolated narrow absorption features in the Ca\,\textsc{ii} K \& H and Na\,\textsc{i} D1 \& D2 lines of HIP\,30414 (in black). The known velocities of clouds in the line of sight are plotted in grey. Comparison stars with their narrow absorption features are plotted in various colours, with their distances and angular separations from HIP\,30414 stated. The pink region corresponds to the radial velocity of HIP\,30414, 20.88\,$\pm$\,1.85\,km\,s$^{-1}$. \textbf{Right: }Same but for epoch 2 of HIP\,30414.}
\end{figure*}

\subsection{Spectral analysis of gas features}\label{sec:SpAnalysis}

Our spectral analysis is aimed at identifying new young debris disc systems that exhibit gas detectable through circumstellar absorption. We examined the Ca\,\textsc{ii} K \& H, and Na\,\textsc{i} D1 \& D2 lines in the spectra of all 130 targets described in Section\,\ref{sec:Sample}. Although most discs in this sample show low fractional IR excesses, their young ages make the primordial gas origin possible, although secondary origin cannot be ruled out. 

To determine whether each feature originated from circumstellar or interstellar gas, we implemented methods described in Section\,\ref{sec:MethodsCharacterisation}. Reliable comparison between nearby stars required both similar distances and small angular separations, which are not always possible. Hence, where available, we used multiple comparison stars, since when multiple objects showed absorption at the same velocity, the likelihood of interstellar absorption increased significantly. Features that did not have such matches, remained stable over time, and were at the stellar radial velocity were considered likely circumstellar \citep{Kiefer2014}. 

Across the sample, two stars showed absorption features consistent with circumstellar gas: HIP\,30414 and TYC\,7879-1373-1. In the case of TYC\,6822-283-1, the origin is inconclusive. The parameters of these three systems are shown in Table\,\ref{tab:DiscussedObjectsParams}. HIP\,61782 (also known as HD\,110058), which is also part of our sample, was previously confirmed to host circumstellar gas \citep{Hales2017,Iglesias2018,Rebollido2018TheDiscs}, hence we do not discuss it in detail here. Nevertheless, it is included in our sample for completeness, as our goal was to present the most comprehensive sample to date of all young intermediate-mass stars with 12\,$\mu$m excess emission below protoplanetary disc level \citep{Iglesias2023}.

We summarise the apparent column densities and equivalent widths (EW) of the detected narrow absorption features for the three objects showing possible circumstellar gas in Table\,\ref{tab:ColumnDensity}. Following \citet{Iglesias2018}, the Ca\,\textsc{ii}\,/\,Na\,\textsc{i} apparent column density ratio can provide additional constraints on the nature of the gas, with values $\lesssim$1 commonly associated with interstellar absorption, as the ISM typically contains more neutral sodium relative to ionised calcium due to its weaker radiation field compared to the circumstellar environment \citep{Redfield2002}. However, ratios above 1 are not a definitive indication of circumstellar nature, as some interstellar clouds can also produce high Ca\,\textsc{ii}\,/\,Na\,\textsc{i} values, and hence this ratio cannot be used as a sole diagnostic. The features we classified as interstellar using other methods (detailed in Section\,\ref{sec:MethodsCharacterisation}) generally have Ca\,\textsc{ii}\,/\,Na\,\textsc{i}\,$<$\,1, while features we identified as likely circumstellar have larger Ca\,\textsc{ii}\,/\,Na\,\textsc{i} ratios. This behaviour is consistent with previous studies, but can only be used as a tool in combination with other diagnostic methods.

\subsection{Gas of possible circumstellar origin}\label{sec:ResultsCircum}

\subsubsection{TYC\,7879-1373-1}

The spectrum of TYC\,7879-1373-1 (Figure\,\ref{fig:TYC7879-1373-1_TYC6822-283-1}, left) shows three narrow absorption features in the Ca\,\textsc{ii} K line at 
-11.33\,km\,s$^{-1}$ (7.4$\sigma$), -4.02\,km\,s$^{-1}$ (42.6$\sigma$), and 2.04\,km\,s$^{-1}$ (27.9$\sigma$). We compared it with spectra of four nearby stars - two from our sample and two from the ESO archive. The first absorption feature does not match any known interstellar clouds or absorption features in the spectra of nearby stars, but matches the stellar radial velocity of -10.68\,$\pm$\,1.64\,km\,s$^{-1}$. This feature is absent in the Na\,\textsc{i} D1 and D2 lines, but is prominent in both Ca\,\textsc{ii} K \& H lines. This first narrow gas absorption feature has an apparent column density ratio of (Ca\,\textsc{ii}\,/\,Na\,\textsc{i}) of 3.67 (Table\,\ref{tab:ColumnDensity}), which strengthens its circumstellar nature. We find that TYC\,7879-1373-1 shows no evidence of ongoing accretion.

\subsubsection{HIP\,30414 (HD\,44892): variable circumstellar gas?}\label{sec:HIP30414var}

Most of the variable features we identified in our sample were found in spectroscopic binary systems, where changes between each epoch result from the orbital motion between the component stars. However, one object in our sample, HIP\,30414 (also known as HD\,44892), exhibits variability that cannot be attributed to binarity. As shown in Figure\,\ref{fig:HIP30414}, the spectra of HIP\,30414 change between the two observed epochs. To distinguish stellar absorption from narrow, possibly circumstellar features, we fitted the spectra using Kurucz photospheric models (spectra created using VidmaPy\footnote{\url{https://github.com/RozanskiT/vidmapy}}, models from \citet{CastelliKurucz2003}, see Figure\,\ref{fig:HIP30414_appendix}), which are more accurate than simpler polynomial fits \citep{Iglesias2018}. 

After removing the photospheric contribution, the residual flux of both epochs shows a stable feature, present in all four of the analysed lines, at a radial velocity of 22.47\,km\,s$^{-1}$ (35.7$\sigma$) in the Ca\,\textsc{ii} K line. Although it matches the stellar radial velocity of 20.88\,$\pm$\,1.85\,km\,s$^{-1}$, this feature is also well matched by interstellar clouds and by absorption features seen in the spectra of nearby comparison stars. Hence, we cannot rule out its interstellar nature.  

To quantify the variability of the remaining absorption, we measured equivalent widths independently for each epoch in the analysed Ca\,\textsc{ii} and Na\,\textsc{i} lines. Equivalent widths were calculated over a radial velocity range from $-65$\,km\,s$^{-1}$ to 50\,km\,s$^{-1}$, excluding the stable interstellar feature. The rms noise level ($\sigma$) was estimated using continuum regions outside this range. For the Ca\,\textsc{ii} K line, we found the equivalent widths for epochs 1 and 2 to be 232.38\,m$\si{\angstrom}$ (37$\sigma$) and 229.62\,m$\si{\angstrom}$ (51$\sigma$), respectively. In the Ca\,\textsc{ii} H line, these are 148.26\,m$\si{\angstrom}$ (45$\sigma$) and 138.86\,m$\si{\angstrom}$ (55$\sigma$), for the same epochs. The full results are summarised in Table\,\ref{tab:ColumnDensity}.

\begin{figure*} 
\centering
\includegraphics[width=1.8\columnwidth]{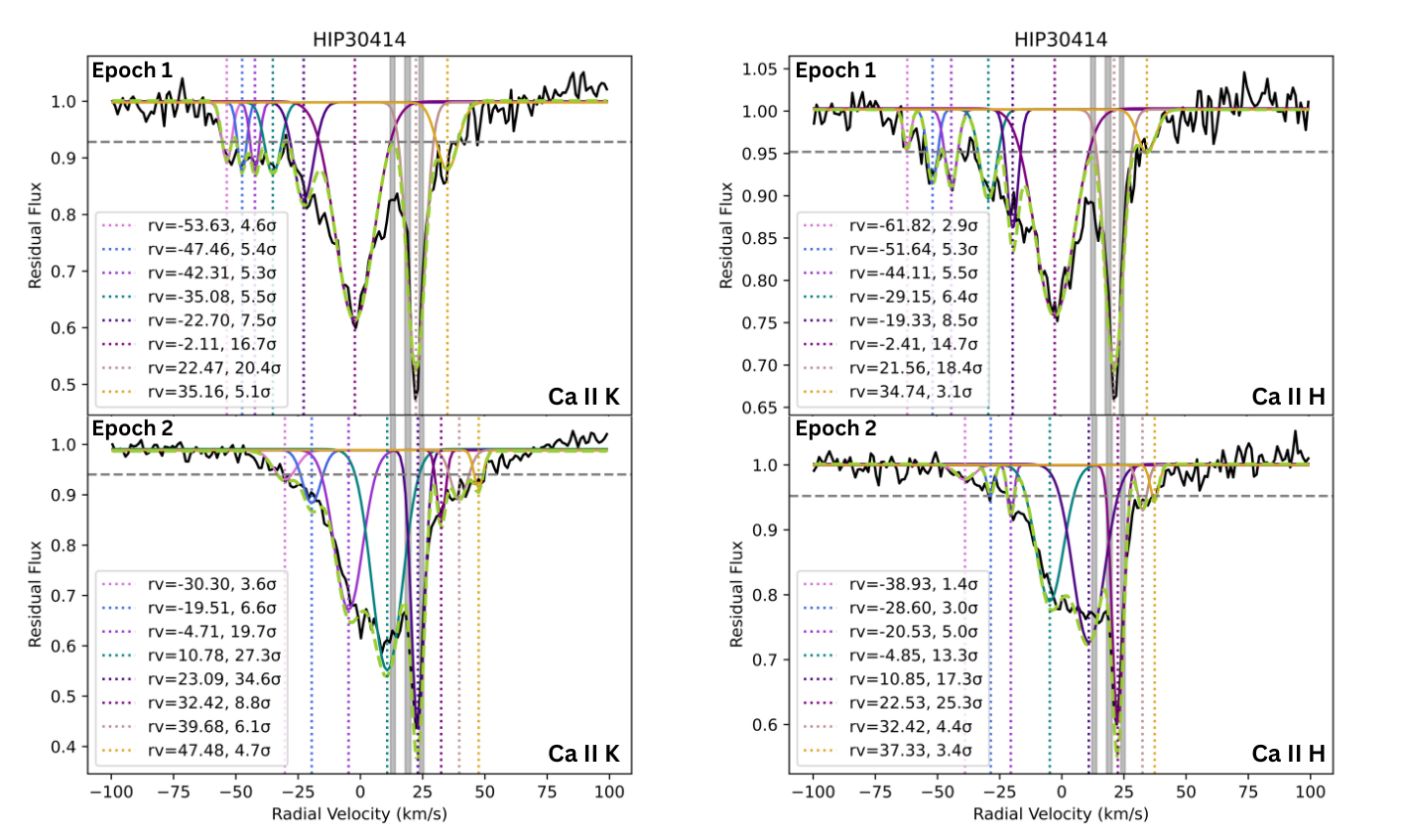}
\caption{\label{fig:HIP30414_variability}Changes in residual flux in the Ca\,\textsc{ii} K (left) and Ca\,\textsc{ii} H (right) lines of HIP\,30414 across two epochs. The residual flux in each epoch is fitted with Gaussian curves, the position and intensities of which vary between epochs 1 and 2. The composite Gaussian fit is shown as a green dashed line, while the residual flux is plotted in black. The positions and intensities of individual Gaussian components are marked accordingly.}
\end{figure*}

Since the equivalent widths between the two epochs are very similar, we chose to showcase the variability by fitting Gaussian curves to Ca\,\textsc{ii} K \& H lines. As shown in Figure\,\ref{fig:HIP30414_variability}, both the positions of the Gaussian peaks and their intensity needed to fit the variable residual absorption change between the two epochs. In the 2nd epoch, significant absorption begins at $\sim$-35\,km\,s$^{-1}$, while in the 1st epoch it extends to $\sim$-55\,km\,s$^{-1}$. The position of the main component also shifts between the two epochs from -2.11\,km\,s$^{-1}$ to 10.78\,km\,s$^{-1}$ in Ca\,\textsc{ii} K and -2.41\,km\,s$^{-1}$ to 10.85\,km\,s$^{-1}$ in Ca\,\textsc{ii} H. The depth of this component changes by 13.4\% and 12.2\% between the two epochs in  Ca\,\textsc{ii} K \& H, respectively.

For the variable features of HIP\,30414, the apparent column density (Ca\,\textsc{ii}\,/\,Na\,\textsc{i}) ratios (Table\,\ref{tab:ColumnDensity}) are quite high ($\sim$\,2.8) whereas the ratios for the stable feature at $\sim$\,22\,km\,s$^{-1}$ are much lower ($\sim$\,0.6). While these ratios alone are not sufficient to determine the nature of gas, their significant difference supports an interstellar origin for the stable component and a circumstellar nature for the variable absorption. 

\begin{figure*} 
    \centering
    \includegraphics[width=1.9\columnwidth]{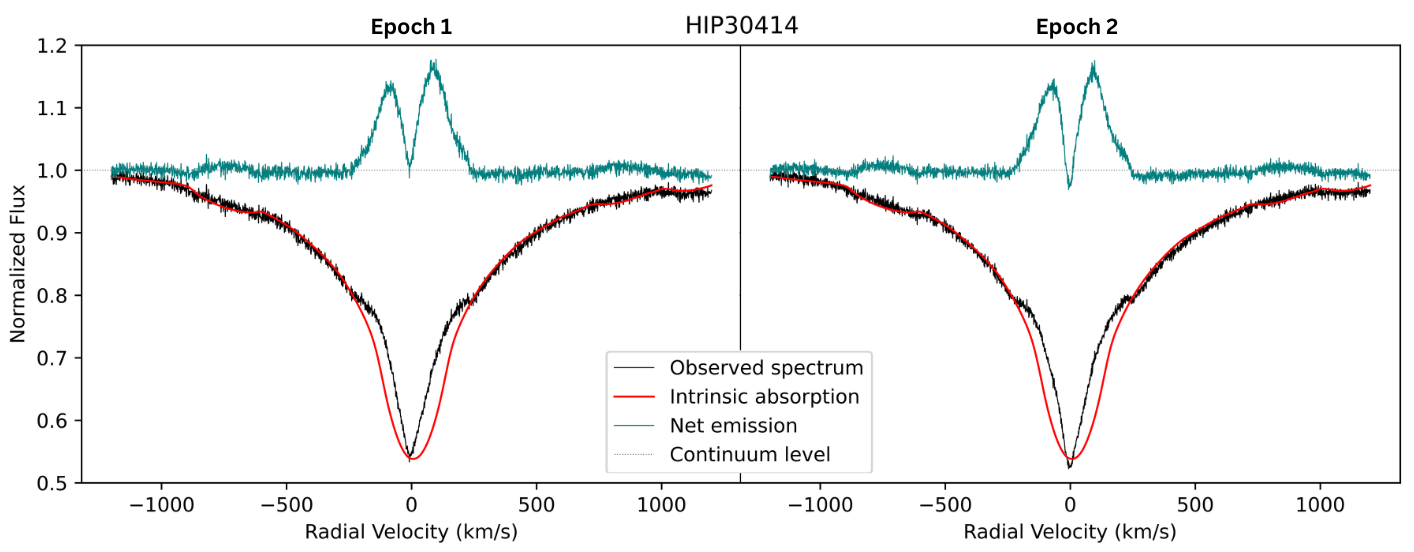}
    \caption{Changes in the H$_{\alpha}$ features across the two observed epochs of HIP\,30414. Left panel shows epoch 1, and the right panel shows epoch 2. The observed flux is plotted in black, while the Kurucz model used for this fit is shown in red (parameters: temperature 7000\,K, log(g) 3.6, [Fe/H] 0 and vsin(i) 140\,km\,s$^{-1}$). The residual (net) emission, shown in teal, contains both emission and absorption features. In epoch 2, the central absorption extends below the continuum level.}
    \label{fig:Halpha}
\end{figure*}

HIP\,30414 is also the only object in our sample exhibiting variability in the H$_{\alpha}$ line (see Figure\,\ref{fig:Halpha}). The residual H$_{\alpha}$ spectrum, obtained by normalizing the observed flux by the fitted Kurucz model, is a composite of emission and absorption components. The relative strength of absorption and emission differs between the two epochs; in epoch 2, the central absorption dominates over emission and extends below the continuum level. Using methods described in \citet{Fairlamb2017ALines}, we estimate the mass accretion rates in the two epochs as $\log\dot{M}_{acc}$\,=\,-7.64$^{+0.11}_{-0.07}\,M_{\odot}\,yr^{-1}$ and $\log\dot{M}_{acc}$\,=\,-7.69$^{+0.11}_{-0.07}\,M_{\odot}\,yr^{-1}$.

\subsection{Inconclusive gas origin}

\subsubsection{TYC\,6822-283-1}

The spectrum of TYC\,6822-283-1 has multiple absorption features in the Ca\,\textsc{ii} K line (Figure\,\ref{fig:TYC7879-1373-1_TYC6822-283-1}, right), all of which are above the noise threshold ($3\sigma$) and were best fitted with a combination of five Gaussian functions. The last two peaks in Ca\,\textsc{ii} K, at 7.16\,km\,s$^{-1}$ (36.5$\sigma$) and 12.29\,km\,s$^{-1}$ (10.7$\sigma$), do not match with the spectra of nearby stars or interstellar clouds, nor do they match the stellar radial velocity. Therefore, with the current observations, we do not have enough information to classify the origin of these features. Their apparent column density ratio (Table\,\ref{tab:ColumnDensity}) are high ($>$\,4), yet this diagnostic alone, as discussed in Section\,\ref{sec:SpAnalysis}, is not enough to conclusively determine the nature of the gas. Additionally, we find no evidence of ongoing accretion in the spectrum of TYC\,6822-283-1.

\subsection{Gas of interstellar origin}

The majority of the narrow absorption features we detected in our sample are consistent with an interstellar gas origin. These features either matched the radial velocities of known interstellar clouds, features found in the spectra of nearby stars, or were detected at velocities not consistent with the stellar radial velocities reported by \citet{Iglesias2023}. Based on these criteria, we classify 111 targets as having absorption features likely interstellar in origin.

In some cases, the depths of absorption-like features were close to our 3$\sigma$ detection threshold, hence we examined their presence across individual epochs. If a feature persisted at the same radial velocity for all epochs and exceeded 3$\sigma$, we considered it to be a detection; if it was not consistent, we deemed it unreliable and thought of it as noise. Across the full sample, we found 15 targets that showed no reliable narrow absorption features.

HIP\,48613 is an example of a target  we ultimately classified as having interstellar gas absorption. The system exhibits a narrow absorption feature close to the stellar radial velocity (feature: 16.46\,km\,s$^{-1}$, stellar: 15\,$\pm$\,8\,km\,s$^{-1}$) that did not correspond to any known interstellar clouds and was not fully reproduced in the spectra of comparison stars. However, HIP\,48613 is a spectroscopic binary, and while its stellar spectrum varied between the observed epochs due to orbital motion, three narrow absorption features remained constant across epochs. Such behaviour is unexpected for absorption arising in a circumstellar disc around either individual star, as these features would be expected to shift with the host star. Although a circumbinary disc could in principle produce such stationary absorption, these features are also strongly present in Na\,\textsc{i} D1 \& D2 lines and matched by comparison stars in sodium absorption. Given that Na\,\textsc{i} is a tracer of interstellar gas \citep{Iglesias2018}, this strongly supports an interstellar origin for the absorption features in HIP\,48613.

\section{Discussion}

\subsection{Identification of circumstellar gas}

We searched for circumstellar gas by identifying narrow absorption lines in stellar spectra, which indicate the presence of gas along the line of sight to the star. Detecting such features is inherently biased towards systems viewed close to edge on, where the observer's line of sight passes through the circumstellar gas disc \citep{Hughes2018,Hales2017}. The two gas-bearing debris disc systems we identified within our sample, TYC\,7879-1373-1 and HIP\,30414, are very young (<\,5\,Myr), and are classified as pre-main-sequence stars - likely caught in the transition from protoplanetary to debris discs, with gas that may still be primordial \citep{Nakatani2021PhotoevaporationRemnants}. Their masses (2.66 and 2.87\,M$_{\odot}$, respectively) place them in the intermediate mass regime, where most circumstellar gas detections have been reported \citep{Wyatt2015}. 

\begin{figure*} 
    \centering
    \includegraphics[width=145mm]{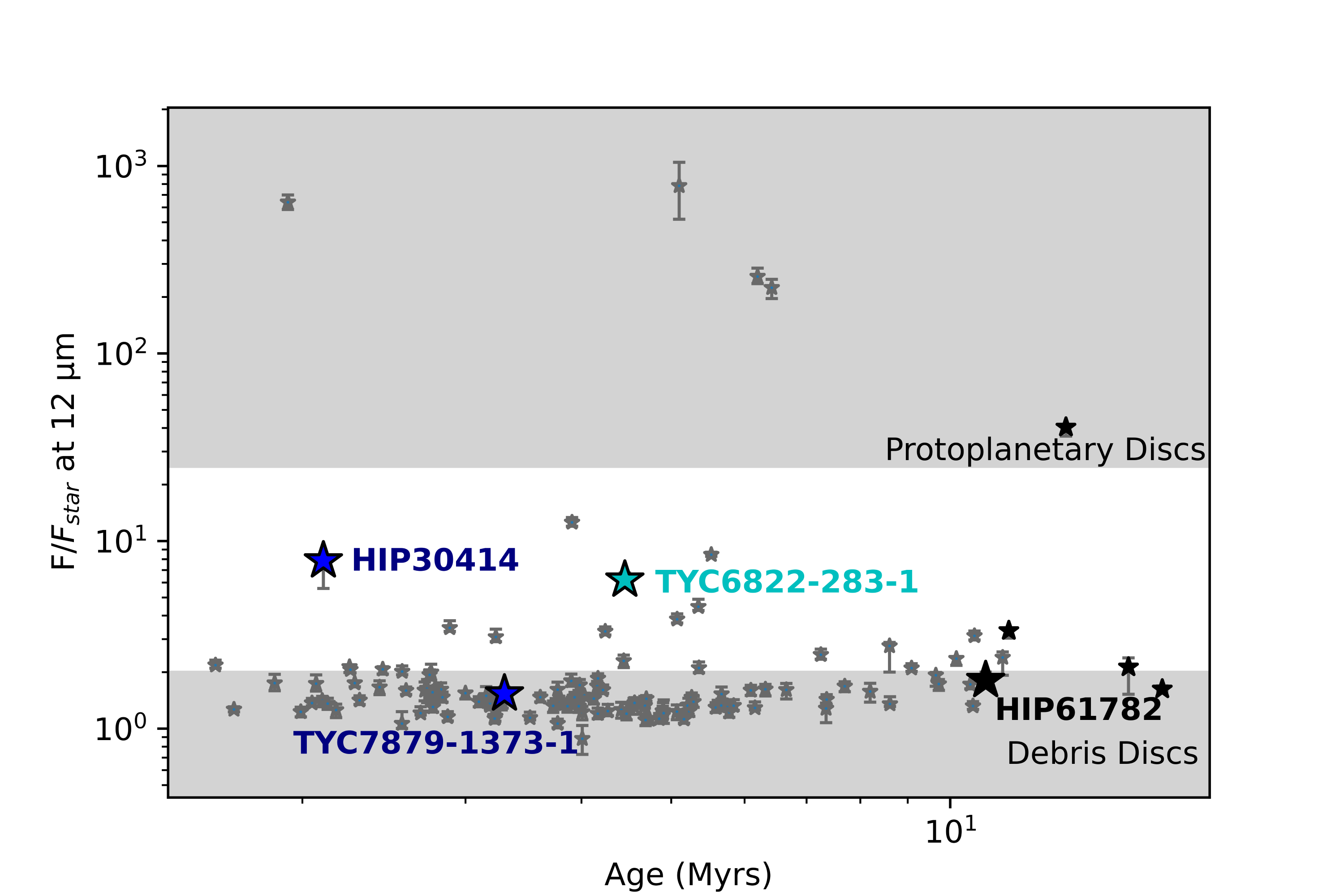}
    \caption{Excess emission at 12\,$\mu$m vs age of 135 identified objects in \citet{Iglesias2023}, plotted with excess uncertainties. Debris and protoplanetary disc regions are shaded. The identified gas-bearing discs marked, including HIP\,61782 not discussed in detail here. The discs caught between protoplanetary and debris disc regimes are: HIP\,30414 and TYC\,7879-1373-1, marked in blue. TYC\,6822-283-1, for which the origin is inconclusive, is marked in cyan. Stars marked in black indicate the objects that may already be on the main-sequence.}
    \label{fig:HybridCandidates}
\end{figure*}

Figure\,\ref{fig:HybridCandidates} presents the ages and 12\,$\mu$m fractional excesses of the 135 targets from \citet{Iglesias2023}, including the five protoplanetary discs (PPDs) we did not analyse. HIP\,30414 occupies the region between protoplanetary and debris discs, while TYC\,7879-1373-1 is placed firmly in the debris category by the level of 12\,$\mu$m fractional excess. TYC\,6822-283-1, for which the origin of gas remains inconclusive, and HIP\,61782, in which gas has been detected by multiple past studies \citep{Hales2017,Iglesias2018,Rebollido2018TheDiscs}, are also shown for context. In the following sections we examine in detail the gas absorption features detected towards these systems (excluding HIP\,61782). Where sufficient data is available, we compare our results with previous studies to provide additional insight and place our findings in a broader context.

\subsection{Debris discs with circumstellar gas}

\subsubsection{TYC\,7879-1373-1}

\citet{Iglesias2023} classify TYC\,7879-1373-1 as a debris disc; however, there are no other mentions of this object in the literature relevant to our study. It exhibits a narrow absorption feature suspected to be of circumstellar origin, as it closely matches the stellar radial velocity. This feature does not correspond to absorption in nearby comparison stars or known interstellar clouds, and is also absent from Na\,\textsc{i} lines, supporting its circumstellar nature. 

Given its young age, TYC\,7879-1373-1 is a strong candidate for hosting a gas-bearing debris disc and an important system for understanding the transition from protoplanetary to debris discs. Further observations, such as gas emission studies, are needed to characterise the gas disc in more detail. Nonetheless, based on the available data, we consider this detection to originate from circumstellar gas. 

\subsubsection{HIP\,30414 (HD\,44892): variable circumstellar gas}

Based on the H$_{\alpha}$ line, we found that the pre-MS star HIP\,30414 is undergoing active accretion, which makes the inner disc a strong candidate for retaining primordial circumstellar gas. Evidence for a dusty circumstellar disc was first suggested by \citet{Patten1991}, who reported the presence of 12\,$\mu$m and 25\,$\mu$m excesses detected with IRAS. \citet{Ishihara2017FaintIRSF} also identified it as a debris disc candidate based on the level of IR excess emission found at 18\,$\mu$m, while \citet{Iglesias2023} first report it as a 'hybrid disc candidate' due to its relatively high 12\,$\mu$m excess.

The residual spectra of HIP\,30414 show variability that is absent from the spectra of nearby stars in the corresponding radial velocity range. Similarly, no known interstellar clouds are present in this velocity range. The stable feature at $\sim$22\,km\,s$^{-1}$ is well matched by the best available comparison star, HIP\,28877, in all lines including sodium, confirming its interstellar origin. However, the variable absorption component, that is seen in both Ca\,\textsc{ii} lines, is not detected in Na\,\textsc{i} lines and does not correspond to any known interstellar clouds. This strongly suggests that the variable feature arises from transient circumstellar gas, like exocomets \citep[like in $\beta$\,Pic][]{Dent2014MolecularDisk}, a wind disc \citep[like in NO\,Lup][]{Lovell2021RapidDisc}, or accretion.

The correlation in variability in the Ca\,\textsc{II} lines and the H$_{\alpha}$ line indicates that the gas is hot and located close to the star, and likely primordial rather than secondary in origin, as secondary gas would be unlikely to sustain the high accretion rate we derived (see Section\,\ref{sec:HIP30414var}). Given that HIP\,30414 is actively accreting, the variability we observed is most likely associated with the ongoing accretion rather than exocomets or winds. Furthermore, as the gas absorption in the H$_{\alpha}$ line overcomes much of the gas emission, the system is likely seen edge-on.

Further supporting the presence of circumstellar gas in HIP\,30414 is the recent discovery of $^{12}$CO (J\,=\,2\,-\,1) and continuum emission with ALMA in this debris disc in \citet{Szewczyk2025}. The reported CO gas mass in this disc is between 10$^{-4}$\,M$_{\oplus}$ and 10$^{-2}$\,M$_{\oplus}$. This detection, combined with the evidence of ongoing gas accretion, suggests that HIP\,30414 may be in the rare transitional phase between protoplanetary and debris disc stages.

\subsection{Inconclusive gas origin}

\subsubsection{TYC\,6822-283-1}

TYC\,6822-283-1 was first reported as a 'hybrid disc candidate' by \citet{Iglesias2023}, who also first identified its IR excess levels in the WISE W1-W4 bands, between protoplanetary and debris disc levels. We found that the spectrum of TYC\,6822-283-1 exhibits two pronounced absorptions with no clear origin (Figure\,\ref{fig:TYC7879-1373-1_TYC6822-283-1}, right). These features do not correspond to known interstellar clouds or nearby stars, nor do they precisely match the stellar radial velocity. This object is located in a region where multiple interstellar clouds are present, complicating the classification. While it could originate from circumstellar gas, interstellar origin cannot be ruled out. 

On the age vs 12\,$\mu$m fractional excess plot (Figure\,\ref{fig:HybridCandidates}), TYC\,6822-283-1 has a relatively high level of IR excess, suggesting the presence of substantial dust in the disc. Its estimated age of 4\,Myr \citep{Iglesias2023} classifies it as a pre-main-sequence star, increasing the likelihood of retaining detectable primordial gas. We searched the literature and found no relevant studies that could provide insight into the origin of these features, hence it remains inconclusive until we are able to collect further evidence.  

\section{Conclusions}

We analysed the spectra of 130 pre-main-sequence debris discs and 'hybrid disc candidates' from \citet{Iglesias2023}, searching for narrow absorption features in Ca\,\textsc{ii} K \& H and Na\,\textsc{i} D1 \& D2 lines that could indicate the presence of circumstellar gas. Absorption spectroscopy is most effective for identifying gas along the line of sight, favouring systems where the disc is viewed edge-on. Because of this observational bias, the likelihood of interstellar origin of the gas absorption lines is high, hence we employed several diagnostic methods to determine the nature of the features. This included comparison to nearby stars, searching for variability between observed epochs, calculating line ratios, and others (see Section\,\ref{sec:MethodsCharacterisation} for details), to distinguish between circumstellar and interstellar nature.  

Through this process, we identified two new systems with likely circumstellar gas: TYC\,7879-1373-1, which is a new detection, and HIP\,30414, in which CO gas emission was previously identified by \citet{Szewczyk2025}. We also confirm the presence of a gas disc in HIP\,61782, where circumstellar absorption was previously reported by \citet{Hales2017}, \citet{Iglesias2018}, and \citet{Rebollido2018TheDiscs}, hence we do not discuss this system in detail. In addition, one system, TYC\,6822-283-1, shows narrow absorption features for which the origin of gas remains inconclusive based on our diagnostic methods. Out of the remaining targets, 111 systems show features that are interstellar in origin, while 15 candidates exhibit no detectable narrow absorption features in the analysed lines. Our results are summarised in Table\,\ref{tab:AllObjects}. Despite the small number of circumstellar detections, our results show that absorption spectroscopy is a time-effective screening tool when searching for gas-bearing discs within large target samples.

The two new systems with likely circumstellar gas are both pre-MS stars (<\,5\,Myr), with stellar masses of 2.66\,M$_{\odot}$ (TYC\,7879-1373-1) and 2.87\,M$_{\odot}$ (HIP\,30414), placing them in the intermediate mass regime where circumstellar gas is most commonly detected \citep{Wyatt2015, Hughes2018}. Their young ages suggest that the gas in them may still be of primordial origin \citep{Nakatani2021PhotoevaporationRemnants,Smirnov-Pinchukov2022LackGas}, although we cannot rule out the secondary origin. HIP\,30414, in particular, shows evidence of ongoing active accretion based on the variability in H$\alpha$ emission we have measured. This provides further evidence for presence of primordial gas located close to the star. TYC\,6822-283-1 is also a very young (4\,Myr) intermediate mass star (2.50\,M$_{\odot}$), however, until we are able to collect more data, we cannot rule out the possibility of interstellar nature of the detected gas. 

Assuming the typical vertical aspect ratio of debris discs as 0.07, taken as a median of values from the REASONS study \citep{Matra2025REASONS}, a typical angle such a disc spawns is 4$\degree$. For a uniform distribution of sin(i) between 0 and 1, the probability of $90^\degree-4^\degree\,\le i\,\le\,90^\degree+4^\degree$ (i.e., near edge-on disc) is $\sim$7\% (corresponding to $\sim$9 discs in our sample). Out of 130 systems, we detect likely circumstellar features in three discs, including HIP\,61782 \citep[which is confirmed to be a near edge-on disc 
in][]{Hales2022}. This corresponds to $\sim$33\% of discs observed near edge-on, considering a uniform distribution of inclinations. While this suggests that around a third of edge-on debris discs in our sample may exhibit detectable gas absorption, the detection number is small and hence this estimate is highly uncertain and should not be considered as representative of the full debris disc population. Nevertheless, our full sample is not located in one place in the sky or limited to star-forming regions, reducing observational bias. 

While the circumstellar origin of their features cannot be certain, additional studies can help determine their nature. Future observations can confirm the presence of gas-bearing discs in the identified systems. For TYC\,7879-1373-1, ALMA observations targetting continuum and CO line emission could reveal the dust and gas morphology. HIP\,30414 has already been observed as part of the ALMA programme 2022.1.01686.S (PI Panić), with $^{12}$CO gas and dust emission detected in a Keplerian disc \citep{Szewczyk2025}. TYC\,6822-283-1 was also scheduled for ALMA observations under the same programme, however these were never completed. Continued multi-wavelength studies, including high-resolution spectroscopy and interferometry, will be crucial to better understand the transition from protoplanetary to debris discs, and how gas affects planetary systems' evolution. 

\section*{Acknowledgements}

KS acknowledges funding from the Bell Burnell Graduate Scholarship Fund (BB0023). DPI and OP acknowledge support from the Science and Technology Facilities Council via grant number ST/X001016/1.

This work has made use of data from the European Space Agency (ESA) mission
{\it Gaia} (\url{https://www.cosmos.esa.int/gaia}), processed by the {\it Gaia}
Data Processing and Analysis Consortium (DPAC,
\url{https://www.cosmos.esa.int/web/gaia/dpac/consortium}). Funding for the DPAC has been provided by national institutions, in particular the institutions participating in the {\it Gaia} Multilateral Agreement.

Based on observations collected at the European Southern Observatory under ESO programmes: 112.262P.001, 110.23YZ.001, 113.26V6.001, 109.23K8.001, 092.C-0721(A), 073.D-0049(A), 085.A-9027(B), 098.C-0463(B), 0102.A-9008(A), 0103.C-0206(A), 096.A-9024(A), 194.C-0833(D), 099.D-0380(A), 087.C-0012(B), 075.D-0676(B), 095.A-9029(C), 078.D-0080(A), 092.A-9006(A), 076.D-0169(A), 0100.C-0090(B), 178.D-0361(B), 67.D-0133(A), 099.C-0491(A), 078.A-9059(A), 0101.D-0697(A), 098.C-0739(A), 68.C-0548(A), 072.C-0488(E), 087.A-9013(A), 080.C-0712(A), 194.C-0833(H), 076.C-0279(A), 182.D-0356(B), 090.C-0421(A), 099.C-0637(A), 074.C-0135(A), 185.D-0056(E), 072.D-0021(B), 0106.A-9009(A), 085.D-0093(A), 099.A-9022(A), 0106.A-9006(A), 089.D-0129(A), 085.A-9027(A), 089.C-0006(A), 079.A-9009(A), 087.C-0012(A), 183.C-0972(A), 075.C-0637(A), 080.D-0194(A), 185.D-0056(B), 082.C-0446(B), 083.C-0676(A), 194.C-0833(B), 0102.D-0281(A), 097.A-9024(A), 085.A-9027(G), 097.A-9022(A), 0101.A-9016(A), 096.A-9039(A), 076.D-0018(A), 099.A-9009(A), 073.A-9008(A), 106.20Y7.002, 079.B-0856(A), 099.A-9029(A), 179.C-0197(D), 075.D-0760(A), 190.C-0027(A), 083.A-9014(B), 266.D-5655(A), 075.D-0324(A), 105.20L8.002, 081.C-0475(A), 079.A-9014(A), 079.D-0118(A), 077.C-0547(A), 073.C-0337(A), 075.D-0103(A), 087.C-0831(A), 0101.C-0232(C), 088.C-0353(A), 0103.C-0548(A), 097.C-0409(B), 185.D-0056(I), 185.D-0056(C), 091.C-0713(A), 67.D-0579(A), 079.C-0789(A), 0104.A-9010(A), 084.D-0067(A), 0100.C-0097(A), 0102.C-0338(A), 074.D-0008(B), 0104.A-9004(A), 184.C-0815(E), 094.C-0946(A), 082.C-0446(A), 081.C-2003(A), 179.C-0197(C), 0103.A-9009(A), 084.A-9011(B), 075.C-0479(A), 083.A-9013(A), 090.D-0358(A), 185.D-0056(A), 088.A-9029(A), 072.D-0021(A), 097.C-0090(A), 088.D-0026(D), 073.D-0641(A), 0103.A-9011(A), 0102.A-9010(A), 179.C-0197(A), 082.C-0831(A), 073.D-0724(A), 0100.A-9006(A), 077.D-0247(A), 0102.C-0584(A), 087.B-0600(G), 085.C-0019(A), 66.D-0457(A), 105.2045.001, 093.C-0658(A), 0102.C-0547(A), 0104.A-9003(A), 083.D-0034(A), 106.21R4.001, 60.A-9700(A), 077.C-0192(A), 097.C-0409(A), 075.C-0689(B), 097.D-0150(A), 0101.C-0232(A), 095.A-9032(A), 0100.C-0414(B), 71.B-0529(A), 0108.A-9006(A), 087.A-9029(A), 087.D-0950(B), 093.C-0409(A), 0101.A-9001(A), 184.C-0815(A), 0101.D-0415(A), 077.C-0575(A), 096.A-9027(A), 293.D-5036(A), 0101.D-0921(A), 098.C-0463(C), 089.D-0975(A), 184.C-0815(C), 083.C-0139(A), 098.A-9039(C) 380.C-0083(A) 077.C-0295(D), 093.D-0302(A), 096.A-9018(A), 074.D-0021(A), 076.C-0279(B), 079.D-0009(C), 077.C-0295(A), 0104.C-0418(C), 091.D-0221(A), 092.A-9002(A), 0101.A-9005(A), 072.D-0149(A), 077.C-0295(B), 105.2045.002, 087.D-0010(A), 0108.A-9029(A), 092.C-0173(A), 105.2055.001, 075.D-0597(A), 0102.C-0040(B), 096.C-0238(A), 094.A-9012(A), 082.D-0061(A), 69.C-0171(A), 076.C-0279(C), 077.D-0720(B), 184.C-0815(F), 075.C-0689(A), 380.C-0083(B), 165.N-0276(A), 60.A-9700(G), 082.B-0484(A), 0104.A-9001(A), 084.B-0029(A), 079.D-0567(A), 088.D-0026(D), 075.D-0145(A), 095.A-9029(C), 074.D-0240(A), 080.C-0032(B), 079.C-0170(A), 074.C-0102(A), 075.C-0234(A), 082.C-0308(A), 080.C-0032(A), 076.C-0073(A), 095.D-0717(A), 080.D-0191(A), 072.D-0196(A), 077.D-0478(A), and 0103.D-0119(A).

\section*{Data Availability}

All data underlying this article are available in the ESO and GAIA archives. ESO data
are found at
\url{http://archive.eso.org/cms.html}. GAIA archive can be browsed at
\url{https://gea.esac.esa.int/archive}.



\bibliographystyle{mnras}
\bibliography{example} 

@ARTICLE{Hales2017,
       author = {{Hales}, Antonio S. and {Barlow}, M.~J. and {Crawford}, I.~A. and {Casassus}, S.},
        title = "{Atomic gas in debris discs}",
      journal = {\mnras},
         year = 2017,
        month = apr,
       volume = {466},
       number = {3},
        pages = {3582-3593},
        doi = {10.1093/mnras/stw3274},
archivePrefix = {arXiv},
       eprint = {1612.05465},
 primaryClass = {astro-ph.EP},
       adsurl = {https://ui.adsabs.harvard.edu/abs/2017MNRAS.466.3582H}
}

@ARTICLE{Iglesias2018,
       author = {{Iglesias}, D. and {Bayo}, A. and {Olofsson}, J. and {Wahhaj}, Z. and {Eiroa}, C. and {Montesinos}, B. and {Rebollido}, I. and {Smoker}, J. and {Sbordone}, L. and {Schreiber}, M.~R. and {Henning}, Th},
        title = "{Debris discs with multiple absorption features in metallic lines: circumstellar or interstellar origin?}",
      journal = {\mnras},
         year = 2018,
        month = oct,
       volume = {480},
       number = {1},
        pages = {488-520},
        doi = {10.1093/mnras/sty1724},
archivePrefix = {arXiv},
       eprint = {1806.10687},
 primaryClass = {astro-ph.SR},
       adsurl = {https://ui.adsabs.harvard.edu/abs/2018MNRAS.480..488I}
}

@ARTICLE{Hughes2018,
       author = {{Hughes}, A. Meredith and {Duch{\^e}ne}, Gaspard and {Matthews}, Brenda C.},
        title = "{Debris Disks: Structure, Composition, and Variability}",
      journal = {\araa},
         year = 2018,
        month = sep,
       volume = {56},
        pages = {541-591},
        doi = {10.1146/annurev-astro-081817-052035},
archivePrefix = {arXiv},
       eprint = {1802.04313},
 primaryClass = {astro-ph.EP},
       adsurl = {https://ui.adsabs.harvard.edu/abs/2018ARA&A..56..541H}
}

@ARTICLE{Wyatt2015,
       author = {{Wyatt}, M.~C. and {Pani{\'c}}, O. and {Kennedy}, G.~M. and {Matr{\`a}}, L.},
        title = "{Five steps in the evolution from protoplanetary to debris disk}",
      journal = {\apss},
         year = 2015,
        month = jun,
       volume = {357},
       number = {2},
          eid = {103},
        pages = {103},
        doi = {10.1007/s10509-015-2315-6},
archivePrefix = {arXiv},
       eprint = {1412.5598},
 primaryClass = {astro-ph.EP},
       adsurl = {https://ui.adsabs.harvard.edu/abs/2015Ap&SS.357..103W}
}

@ARTICLE{Kiefer2014,
       author = {{Kiefer}, F. and {Lecavelier des Etangs}, A. and {Augereau}, J. -C. and {Vidal-Madjar}, A. and {Lagrange}, A. -M. and {Beust}, H.},
        title = "{Exocomets in the circumstellar gas disk of HD 172555}",
      journal = {\aap},
         year = 2014,
        month = jan,
       volume = {561},
          eid = {L10},
        pages = {L10},
         doi = {10.1051/0004-6361/201323128},
archivePrefix = {arXiv},
       eprint = {1401.1365},
 primaryClass = {astro-ph.EP},
       adsurl = {https://ui.adsabs.harvard.edu/abs/2014A&A...561L..10K}
}

@ARTICLE{Redfield2008,
       author = {{Redfield}, Seth and {Linsky}, Jeffrey L.},
        title = "{The Structure of the Local Interstellar Medium. IV. Dynamics, Morphology, Physical Properties, and Implications of Cloud-Cloud Interactions}",
      journal = {\apj},
         year = 2008,
        month = jan,
       volume = {673},
       number = {1},
        pages = {283-314},
        doi = {10.1086/524002},
archivePrefix = {arXiv},
       eprint = {0804.1802},
 primaryClass = {astro-ph},
       adsurl = {https://ui.adsabs.harvard.edu/abs/2008ApJ...673..283R}
}

@INPROCEEDINGS{TopCat,
       author = {{Taylor}, M.~B.},
        title = "{TOPCAT \& STIL: Starlink Table/VOTable Processing Software}",
    booktitle = {Astronomical Data Analysis Software and Systems XIV},
         year = 2005,
       editor = {{Shopbell}, P. and {Britton}, M. and {Ebert}, R.},
       series = {Astronomical Society of the Pacific Conference Series},
       volume = {347},
        month = dec,
        pages = {29},
}

@article{Simbad,
    author = {{Wenger}, M. and {Ochsenbein}, F. and {Egret}, D. and {Dubois}, P. and {Bonnarel}, F. and {Borde}, S. and {Genova}, F. and {Jasniewicz}, G. and {Lalo{\"e}}, S. and {Lesteven}, S. and {Monier}, R.},
    title = "{The SIMBAD astronomical database. The CDS reference database for astronomical objects}",
      journal = {\aaps},
         year = 2000,
        month = apr,
       volume = {143},
        pages = {9-22},
          doi = {10.1051/aas:2000332},
archivePrefix = {arXiv},
       eprint = {astro-ph/0002110},
 primaryClass = {astro-ph},
       adsurl = {https://ui.adsabs.harvard.edu/abs/2000A&AS..143....9W}
}

@ARTICLE{Kral2017,
       author = {{Kral}, Quentin and {Matr{\`a}}, Luca and {Wyatt}, Mark C. and {Kennedy}, Grant M.},
        title = "{Predictions for the secondary CO, C and O gas content of debris discs from the destruction of volatile-rich planetesimals}",
      journal = {\mnras},
         year = 2017,
        month = jul,
       volume = {469},
       number = {1},
        pages = {521-550},
}

@ARTICLE{Iglesias2023,
       author = {{Iglesias}, Daniela P. and {Pani{\'c}}, Olja and {van den Ancker}, Mario and {Petr-Gotzens}, Monika G. and {Siess}, Lionel and {Vioque}, Miguel and {Pascucci}, Ilaria and {Oudmaijer}, Ren{\'e} and {Miley}, James},
        title = "{X-shooter survey of young intermediate-mass stars - I. Stellar characterization and disc evolution}",
      journal = {\mnras},
         year = 2023,
        month = mar,
       volume = {519},
       number = {3},
        pages = {3958-3975},
         doi = {10.1093/mnras/stac3619},
archivePrefix = {arXiv},
       eprint = {2212.06791},
 primaryClass = {astro-ph.SR},
       adsurl = {https://ui.adsabs.harvard.edu/abs/2023MNRAS.519.3958I}
}

@ARTICLE{Hales2022,
       author = {{Hales}, Antonioand {Marino}, Sebasti{\'a}n and {Sheehan}, Patrick D. and {Ulloa}, Silvio and {P{\'e}rez}, Sebasti{\'a}n and {Matr{\`a}}, Luca and {Kral}, Quentin and {Wyatt}, Mark and {Dent}, William and {Carpenter}, John},
        title = "{ALMA Observations of the HD 110058 Debris Disk}",
      journal = {\apj},
         year = 2022,
        month = dec,
       volume = {940},
       number = {2},
          eid = {161},
        pages = {161},
        doi = {10.3847/1538-4357/ac9cd3},
archivePrefix = {arXiv},
       eprint = {2210.12275},
 primaryClass = {astro-ph.SR},
       adsurl = {https://ui.adsabs.harvard.edu/abs/2022ApJ...940..161H}
}

@INPROCEEDINGS{Dekker2000,
   author = {{Dekker}, H. and {D'Odorico}, S. and {Kaufer}, A. and {Delabre}, B. and 
	{Kotzlowski}, H.},
    title = "{Design, construction, and performance of UVES, the echelle spectrograph for the UT2 Kueyen Telescope at the ESO Paranal Observatory}",
booktitle = {Optical and IR Telescope Instrumentation and Detectors},
     year = 2000,
   series = {\procspie},
   volume = 4008,
   editor = {{Iye}, M. and {Moorwood}, A.~F.},
    month = aug,
    pages = {534-545},
   doi = {10.1117/12.395512},
       adsurl = {https://ui.adsabs.harvard.edu/abs/2000SPIE.4008..534D}
}

@ARTICLE{Freudling2013,
   author = {{Freudling}, W. and {Romaniello}, M. and {Bramich}, D.~M. and
{Ballester}, P. and {Forchi}, V. and {Garc{\'{\i}}a-Dabl{\'o}}, C.~E. and
{Moehler}, S. and {Neeser}, M.~J.},
    title = "{Automated data reduction workflows for astronomy. The ESO Reflex environment}",
  journal = {\aap},
archivePrefix = "arXiv",
   eprint = {1311.5411},
 primaryClass = "astro-ph.IM",
     year = 2013,
    month = nov,
   volume = 559,
      eid = {A96},
    pages = {A96},
      doi = {10.1051/0004-6361/201322494},
   adsurl = {http://adsabs.harvard.edu/abs/2013A%26A...559A..96F}
}

@article{Nakatani2021PhotoevaporationRemnants,
    title = {{Photoevaporation of Grain-depleted Protoplanetary Disks around Intermediate-mass Stars: Investigating the Possibility of Gas-rich Debris Disks as Protoplanetary Remnants}},
    year = {2021},
    journal = {The Astrophysical Journal},
    author = {Nakatani, Riouhei and Kobayashi, Hiroshi and Kuiper, Rolf and Nomura, Hideko and Aikawa, Yuri},
    number = {2},
    month = {7},
    pages = {90},
    volume = {915},
    doi = {10.3847/1538-4357/ac0137},
    issn = {0004-637X}
}

@article{Kral2019ImagingDiscs,
    title = {{Imaging [CI] around HD 131835: reinterpreting young debris discs with protoplanetary disc levels of CO gas as shielded secondary discs}},
    year = {2019},
    journal = {Monthly Notices of the Royal Astronomical Society},
    author = {Kral, Quentin and Marino, Sebastian and Wyatt, Mark C and Kama, Mihkel and Matr{\'{a}}, Luca},
    number = {3},
    month = {11},
    pages = {3670--3691},
    volume = {489},
    doi = {10.1093/mnras/sty2923},
    issn = {0035-8711}
}

@article{Marino2022VerticalLifetime,
    title = {{Vertical evolution of exocometary gas – I. How vertical diffusion shortens the CO lifetime}},
    year = {2022},
    journal = {Monthly Notices of the Royal Astronomical Society},
    author = {Marino, S and Cataldi, G and Jankovic, M R and Matr{\`{a}}, L and Wyatt, M C},
    number = {1},
    month = {7},
    pages = {507--524},
    volume = {515},
    doi = {10.1093/mnras/stac1756},
    issn = {0035-8711}
}

@article{Lovell2021RapidDisc,
    title = {{Rapid CO gas dispersal from NO Lup’s class III circumstellar disc}},
    year = {2021},
    journal = {Monthly Notices of the Royal Astronomical Society: Letters},
    author = {Lovell, J B and Kennedy, G M and Marino, S and Wyatt, M C and Ansdell, M and Kama, M and Manara, C F and Matr{\`{a}}, L and Rosotti, G and Tazzari, M and Testi, L and Williams, J P},
    number = {1},
    month = {1},
    pages = {L66-L71},
    volume = {502},
    doi = {10.1093/mnrasl/slaa189},
    issn = {1745-3925}
}

@article{Pericaud2017TheDisks,
    title = {{The hybrid disks: a search and study to better understand evolution of disks}},
    year = {2017},
    journal = {Astronomy {\&} Astrophysics},
    author = {P{\'{e}}ricaud, J. and Di Folco, E. and Dutrey, A. and Guilloteau, S. and Pi{\'{e}}tu, V.},
    month = {4},
    pages = {A62},
    volume = {600},
    doi = {10.1051/0004-6361/201629371},
    issn = {0004-6361}
}

@article{Kral2020Survey129590,
    title = {{Survey of planetesimal belts with ALMA: gas detected around the Sun-like star HD 129590}},
    year = {2020},
    journal = {Monthly Notices of the Royal Astronomical Society},
    author = {Kral, Quentin and Matr{\`{a}}, Luca and Kennedy, Grant M and Marino, Sebastian and Wyatt, Mark C},
    number = {3},
    month = {9},
    pages = {2811--2830},
    volume = {497},
    doi = {10.1093/mnras/staa2038},
    issn = {0035-8711}
}

@article{Lovell2021ALMADispersal,
    title = {{ALMA survey of Lupus class III stars: Early planetesimal belt formation and rapid disc dispersal}},
    year = {2021},
    journal = {Monthly Notices of the Royal Astronomical Society},
    author = {Lovell, J B and Wyatt, M C and Ansdell, M and Kama, M and Kennedy, G M and Manara, C F and Marino, S and Matr{\`{a}}, L and Rosotti, G and Tazzari, M and Testi, L and Williams, J P},
    number = {4},
    month = {1},
    pages = {4878--4900},
    volume = {500},
    doi = {10.1093/mnras/staa3335},
    issn = {0035-8711}
}

@article{Lieman-Sifry2016DebrisALMA,
    title = {{Debris Disks in the Scorpius-Centaurus OB Association Resolved by ALMA}},
    year = {2016},
    journal = {The Astrophysical Journal},
    author = {Lieman-Sifry, Jesse and Hughes, A. Meredith and Carpenter, John M. and Gorti, Uma and Hales, Antonio and Flaherty, Kevin M.},
    number = {1},
    month = {9},
    pages = {25},
    volume = {828},
    doi = {10.3847/0004-637X/828/1/25},
    issn = {0004-637X}
}

@article{Welsh2015TheAbsorption,
    title = {{The Appearance and Disappearance of Exocomet Gas Absorption}},
    year = {2015},
    journal = {Advances in Astronomy},
    author = {Welsh, Barry Y. and Montgomery, Sharon L.},
    pages = {1--20},
    volume = {2015},
    doi = {10.1155/2015/980323},
    issn = {1687-7969}
}

@article{Matra2017DetectionComets,
    title = {{Detection of Exocometary CO within the 440 Myr Old Fomalhaut Belt: A Similar CO+CO <sub>2</sub> Ice Abundance in Exocomets and Solar System Comets}},
    year = {2017},
    journal = {The Astrophysical Journal},
    author = {Matr{\`{a}}, L. and MacGregor, M. A. and Kalas, P. and Wyatt, M. C. and Kennedy, G. M. and Wilner, D. J. and Duchene, G. and Hughes, A. M. and Pan, M. and Shannon, A. and Clampin, M. and Fitzgerald, M. P. and Graham, J. R. and Holland, W. S. and Pani{\'{c}}, O. and Su, K. Y. L.},
    number = {1},
    month = {6},
    pages = {9},
    volume = {842},
    doi = {10.3847/1538-4357/aa71b4},
    issn = {0004-637X}
}

@ARTICLE{Smirnov-Pinchukov2022LackGas,
       author = {{Smirnov-Pinchukov}, Grigorii V. and {Mo{\'o}r}, Attila and {Semenov}, Dmitry A. and {{\'A}brah{\'a}m}, P{\'e}ter and {Henning}, Thomas and {K{\'o}sp{\'a}l}, {\'A}gnes and {Hughes}, A. Meredith and {di Folco}, Emmanuel},
        title = "{Lack of other molecules in CO-rich debris discs: is it primordial or secondary gas?}",
      journal = {\mnras},
         year = 2022,
        month = feb,
       volume = {510},
       number = {1},
        pages = {1148-1162},
          doi = {10.1093/mnras/stab3146},
archivePrefix = {arXiv},
       eprint = {2111.07655},
 primaryClass = {astro-ph.EP},
       adsurl = {https://ui.adsabs.harvard.edu/abs/2022MNRAS.510.1148S}
}

@article{Miley2018Unlocking141569,
    title = {{Unlocking the secrets of the midplane gas and dust distribution in the young hybrid disc HD 141569}},
    year = {2018},
    journal = {Astronomy {\&} Astrophysics},
    author = {Miley, J. M. and Pani{\'{c}}, O. and Wyatt, M. and Kennedy, G. M.},
    month = {7},
    pages = {L10},
    volume = {615},
    doi = {10.1051/0004-6361/201833381},
    issn = {0004-6361}
}

@article{Rebollido2018TheDiscs,
    title = {{The co-existence of hot and cold gas in debris discs}},
    year = {2018},
    journal = {Astronomy {\&} Astrophysics},
    author = {Rebollido, I. and Eiroa, C. and Montesinos, B. and Maldonado, J. and Villaver, E. and Absil, O. and Bayo, A. and Canovas, H. and Carmona, A. and Chen, Ch. and Ertel, S. and Garufi, A. and Henning, Th. and Iglesias, D. P. and Launhardt, R. and Liseau, R. and Meeus, G. and Mo{\'{o}}r, A. and Mora, A. and Olofsson, J. and Rauw, G. and Riviere-Marichalar, P.},
    month = {6},
    pages = {A3},
    volume = {614},
    doi = {10.1051/0004-6361/201732329},
    issn = {0004-6361}
}

@article{Cataldi2023PrimordialALMA,
    title = {{Primordial or Secondary? Testing Models of Debris Disk Gas with ALMA}},
    year = {2023},
    journal = {The Astrophysical Journal},
    author = {Cataldi, Gianni and Aikawa, Yuri and Iwasaki, Kazunari and Marino, Sebastian and Brandeker, Alexis and Hales, Antonio and Henning, Thomas and Higuchi, Aya E. and Hughes, A. Meredith and Janson, Markus and Kral, Quentin and Matr{\`{a}}, Luca and Mo{\'{o}}r, Attila and Olofsson, Göran and Redfield, Seth and Roberge, Aki},
    number = {2},
    month = {7},
    pages = {111},
    volume = {951},
    doi = {10.3847/1538-4357/acd6f3},
    issn = {0004-637X}
}

@ARTICLE{Patten1991,
       author = {{Patten}, B.~M. and {Willson}, L.~A.},
        title = "{An IRAS Survey of Main-Sequence B,A, and F Stars}",
      journal = {\aj},
         year = 1991,
        month = jul,
       volume = {102},
        pages = {323},
          doi = {10.1086/115879},
       adsurl = {https://ui.adsabs.harvard.edu/abs/1991AJ....102..323P}
}

@article{Ishihara2017FaintIRSF,
    title = {{Faint warm debris disks around nearby bright stars explored by AKARI and IRSF}},
    year = {2017},
    journal = {Astronomy {\&} Astrophysics},
    author = {Ishihara, Daisuke and Takeuchi, Nami and Kobayashi, Hiroshi and Nagayama, Takahiro and Kaneda, Hidehiro and Inutsuka, Shu-ichiro and Fujiwara, Hideaki and Onaka, Takashi},
    month = {5},
    pages = {A72},
    volume = {601},
    doi = {10.1051/0004-6361/201526215},
    issn = {0004-6361}
}

@article{Hobbs1985ThePictoris,
    title = {{The gaseous component of the disk around Beta Pictoris}},
    year = {1985},
    journal = {The Astrophysical Journal},
    author = {Hobbs, L. M. and Vidal-Madjar, A. and Ferlet, R. and Albert, C. E. and Gry, C.},
    month = {6},
    pages = {L29},
    volume = {293},
    doi = {10.1086/184485},
    issn = {0004-637X}
}

@article{Dent2014MolecularDisk,
    title = {{Molecular Gas Clumps from the Destruction of Icy Bodies in the {$\beta$} Pictoris Debris Disk}},
    year = {2014},
    journal = {Science},
    author = {Dent, W. R. F. and Wyatt, M. C. and Roberge, A. and Augereau, J.-C. and Casassus, S. and Corder, S. and Greaves, J. S. and de Gregorio-Monsalvo, I. and Hales, A. and Jackson, A. P. and Hughes, A. Meredith and Lagrange, A.-M. and Matthews, B. and Wilner, D.},
    number = {6178},
    month = {3},
    pages = {1490--1492},
    volume = {343},
    doi = {10.1126/science.1248726},
    issn = {0036-8075}
}

@article{Kral2016APictoris,
    title = {{A self-consistent model for the evolution of the gas produced in the debris disc of {$\beta$} Pictoris}},
    year = {2016},
    journal = {Monthly Notices of the Royal Astronomical Society},
    author = {Kral, Q. and Wyatt, M. and Carswell, R. F. and Pringle, J. E. and Matr{\`{a}}, L. and Juh{\'{a}}sz, A.},
    number = {1},
    month = {9},
    pages = {845--858},
    volume = {461},
    doi = {10.1093/mnras/stw1361},
    issn = {0035-8711}
}

@article{Reffert2015PreciseStars,
    title = {{Precise radial velocities of giant stars}},
    year = {2015},
    journal = {Astronomy {\&} Astrophysics},
    author = {Reffert, Sabine and Bergmann, Christoph and Quirrenbach, Andreas and Trifonov, Trifon and K{\"{u}}nstler, Andreas},
    month = {2},
    pages = {A116},
    volume = {574},
    doi = {10.1051/0004-6361/201322360},
    issn = {0004-6361}
}

@article{Moor2017MolecularStars,
    title = {{Molecular Gas in Debris Disks around Young A-type Stars}},
    year = {2017},
    journal = {The Astrophysical Journal},
    author = {Mo{\'{o}}r, Attila and Cur{\'{e}}, Michel and K{\'{o}}sp{\'{a}}l, Ágnes and {\'{A}}brah{\'{a}}m, Péter and Csengeri, Timea and Eiroa, Carlos and Gunawan, Diah and Henning, Thomas and Hughes, A. Meredith and Juh{\'{a}}sz, Attila and Pawellek, Nicole and Wyatt, Mark},
    number = {2},
    month = {11},
    pages = {123},
    volume = {849},
    doi = {10.3847/1538-4357/aa8e4e},
    issn = {1538-4357}
}

@article{Smette2015Molecfit:Correction,
	author = {{Smette}, A. and {Sana}, H. and {Noll}, S. and {Horst}, H. and {Kausch}, W. and {Kimeswenger}, S. and {Barden}, M. and {Szyszka}, C. and {Jones}, A.~M. and {Gallenne}, A. and {Vinther}, J. and {Ballester}, P. and {Taylor}, J.},
        title = "{Molecfit: A general tool for telluric absorption correction. I. Method and application to ESO instruments}",
      journal = {\aap},
         year = 2015,
        month = apr,
       volume = {576},
          eid = {A77},
        pages = {A77},
          doi = {10.1051/0004-6361/201423932},
archivePrefix = {arXiv},
       eprint = {1501.07239},
 primaryClass = {astro-ph.IM},
       adsurl = {https://ui.adsabs.harvard.edu/abs/2015A&A...576A..77S}
}

@article{Kausch2015Molecfit:Correction,
	author = {{Kausch}, W. and {Noll}, S. and {Smette}, A. and {Kimeswenger}, S. and {Barden}, M. and {Szyszka}, C. and {Jones}, A.~M. and {Sana}, H. and {Horst}, H. and {Kerber}, F.},
        title = "{Molecfit: A general tool for telluric absorption correction. II. Quantitative evaluation on ESO-VLT/X-Shooterspectra}",
      journal = {\aap},
         year = 2015,
        month = apr,
       volume = {576},
          eid = {A78},
        pages = {A78},
          doi = {10.1051/0004-6361/201423909},
archivePrefix = {arXiv},
       eprint = {1501.07265},
 primaryClass = {astro-ph.IM},
       adsurl = {https://ui.adsabs.harvard.edu/abs/2015A&A...576A..78K}
}

@article{Kral2020FormationAccretion,
    title = {{Formation of secondary atmospheres on terrestrial planets by late disk accretion}},
    year = {2020},
    journal = {Nature Astronomy},
    author = {Kral, Quentin and Davoult, Jeanne and Charnay, Benjamin},
    number = {8},
    month = {4},
    pages = {769--775},
    volume = {4},
    doi = {10.1038/s41550-020-1050-2},
    issn = {2397-3366}
}

@article{Mayor2003SettingHARPS,
    title = {{Setting New Standards with HARPS}},
    year = {2003},
    journal = {The Messenger},
    author = {Mayor, M and Pepe, F and Queloz, D and Bouchy, F and Rupprecht, G and Lo Curto, G and Avila, G and Benz, W and Bertaux, J -L. and Bonfils, X and Dall, Th. and Dekker, H and Delabre, B and Eckert, W and Fleury, M and Gilliotte, A and Gojak, D and Guzman, J.~C. and Kohler, D and Lizon, J -L. and Longinotti, A and Lovis, C and Megevand, D and Pasquini, L and Reyes, J and Sivan, J -P. and Sosnowska, D and Soto, R and Udry, S and van Kesteren, A and Weber, L and Weilenmann, U},
    month = {12},
    pages = {20--24},
    volume = {114},
       adsurl = {https://ui.adsabs.harvard.edu/abs/2003Msngr.114...20M}
}

@article{Kaufer1999CommissioningLa-Silla.,
    title = {{Commissioning FEROS, the new high-resolution spectrograph at La-Silla.}},
    year = {1999},
    journal = {The Messenger},
    author = {Kaufer, A and Stahl, O and Tubbesing, S and N{\o}rregaard, P and Avila, G and Francois, P and Pasquini, L and Pizzella, A},
    month = {3},
    pages = {8--12},
    volume = {95}
}

@ARTICLE{Marino2017,
       author = {{Marino}, S. and {Wyatt}, M.~C. and {Pani{\'c}}, O. and {Matr{\`a}}, L. and {Kennedy}, G.~M. and {Bonsor}, A. and {Kral}, Q. and {Dent}, W.~R.~F. and {Duchene}, G. and {Wilner}, D. and {Lisse}, C.~M. and {Lestrade}, J. -F. and {Matthews}, B.},
        title = "{ALMA observations of the {\ensuremath{\eta}} Corvi debris disc: inward scattering of CO-rich exocomets by a chain of 3-30 M$_{{\ensuremath{\oplus}}}$ planets?}",
      journal = {\mnras},
         year = 2017,
        month = mar,
       volume = {465},
       number = {3},
        pages = {2595-2615},
          doi = {10.1093/mnras/stw2867},
archivePrefix = {arXiv},
       eprint = {1611.01168},
 primaryClass = {astro-ph.EP},
       adsurl = {https://ui.adsabs.harvard.edu/abs/2017MNRAS.465.2595M}
}

@ARTICLE{Ribas2015,
       author = {{Ribas}, {\'A}lvaro and {Bouy}, Herv{\'e} and {Mer{\'\i}n}, Bruno},
        title = "{Protoplanetary disk lifetimes vs. stellar mass and possible implications for giant planet populations}",
      journal = {\aap},
         year = 2015,
        month = apr,
       volume = {576},
          eid = {A52},
        pages = {A52},
          doi = {10.1051/0004-6361/201424846},
archivePrefix = {arXiv},
       eprint = {1502.00631},
 primaryClass = {astro-ph.SR},
       adsurl = {https://ui.adsabs.harvard.edu/abs/2015A&A...576A..52R}
}

@article{Wyatt2008EvolutionDisks,
    title = {{Evolution of Debris Disks}},
    year = {2008},
    journal = {Annual Review of Astronomy and Astrophysics},
    author = {Wyatt, Mark C.},
    number = {1},
    month = {9},
    pages = {339--383},
    volume = {46},
    doi = {10.1146/annurev.astro.45.051806.110525},
    issn = {0066-4146}
}

@INPROCEEDINGS{Backman1993,
       author = {{Backman}, Dana E. and {Paresce}, Francesco},
        title = "{Main-Sequence Stars with Circumstellar Solid Material - the VEGA Phenomenon}",
    booktitle = {Protostars and Planets III},
         year = 1993,
       editor = {{Levy}, Eugene H. and {Lunine}, Jonathan I.},
        month = jan,
        pages = {1253},
       adsurl = {https://ui.adsabs.harvard.edu/abs/1993prpl.conf.1253B}
}

@article{Montgomery2012DetectionStars,
    title = {{Detection of Variable Gaseous Absorption Features in the Debris Disks Around Young A-type Stars}},
    year = {2012},
    journal = {Publications of the Astronomical Society of the Pacific},
    author = {Montgomery, Sharon L. and Welsh, Barry Y.},
    number = {920},
    month = {10},
    pages = {1042--1056},
    volume = {124},
    doi = {10.1086/668293},
    issn = {00046280}
}

@ARTICLE{Lagrange-Henri1990SearchStar.,
       author = {{Lagrange-Henri}, A.~M. and {Ferlet}, R. and {Vidal-Madjar}, A. and {Beust}, H. and {Gry}, C. and {Lallement}, R.},
        title = "{Search for beta Pictoris-like star.}",
      journal = {\aaps},
         year = 1990,
        month = nov,
       volume = {85},
        pages = {1089},
       adsurl = {https://ui.adsabs.harvard.edu/abs/1990A&AS...85.1089L}
}

@article{Iglesias2019AnDisc,
    title = {{An unusually large gaseous transit in a debris disc}},
    year = {2019},
    journal = {Monthly Notices of the Royal Astronomical Society},
    author = {Iglesias, Daniela P and Olofsson, Johan and Bayo, Amelia and Zieba, Sebastian and Montesinos, Matías and Smoker, Jonathan and Kennedy, Grant M and Godoy, Nicolás and Pantoja, Blake and Talens, Geert Jan and Wahhaj, Zahed and Zamora, Catalina},
    number = {4},
    month = {12},
    pages = {5218--5227},
    volume = {490},
    doi = {10.1093/mnras/stz2888},
    issn = {0035-8711}
}

@article{Veras2020ConstrainingStars,
    title = {{Constraining planet formation around 6–8 solar mass stars}},
    year = {2020},
    journal = {Monthly Notices of the Royal Astronomical Society},
    author = {Veras, Dimitri and Tremblay, Pier-Emmanuel and Hermes, J J and McDonald, Catriona H and Kennedy, Grant M and Meru, Farzana and G{\"{a}}nsicke, Boris T},
    number = {1},
    month = {3},
    pages = {765--775},
    volume = {493},
    doi = {10.1093/mnras/staa241},
    issn = {0035-8711}
}

@ARTICLE{Bendahan-West2025,
       author = {{Bendahan-West}, Rapha{\"e}l and {Kennedy}, Grant M. and {Brown}, David J.~A. and {Str{\o}m}, Paul A.},
        title = "{Quantifying spectroscopic Ca II exocomet transit occurrence in two decades of HARPS data}",
      journal = {\mnras},
         year = 2025,
        month = feb,
       volume = {537},
       number = {1},
        pages = {229-251},
          doi = {10.1093/mnras/stae2804},
archivePrefix = {arXiv},
       eprint = {2412.13253},
 primaryClass = {astro-ph.EP},
       adsurl = {https://ui.adsabs.harvard.edu/abs/2025MNRAS.537..229B}
}

@article{Fairlamb2017ALines,
	author = {{Fairlamb}, J.~R. and {Oudmaijer}, R.~D. and {Mendigutia}, I. and {Ilee}, J.~D. and {van den Ancker}, M.~E.},
        title = "{A spectroscopic survey of Herbig Ae/Be stars with X-Shooter - II. Accretion diagnostic lines}",
      journal = {\mnras},
         year = 2017,
        month = feb,
       volume = {464},
       number = {4},
        pages = {4721-4735},
          doi = {10.1093/mnras/stw2643},
archivePrefix = {arXiv},
       eprint = {1610.09636},
 primaryClass = {astro-ph.SR},
       adsurl = {https://ui.adsabs.harvard.edu/abs/2017MNRAS.464.4721F}
}

@ARTICLE{Matra2025REASONS,
       author = {{Matr{\`a}}, L. and {Marino}, S. and {Wilner}, D.~J. and {Kennedy}, G.~M. and {Booth}, M. and {Krivov}, A.~V. and {Williams}, J.~P. and {Hughes}, A.~M. and {del Burgo}, C. and {Carpenter}, J. and {Davies}, C.~L. and {Ertel}, S. and {Kral}, Q. and {Lestrade}, J. -F. and {Marshall}, J.~P. and {Milli}, J. and {{\"O}berg}, K.~I. and {Pawellek}, N. and {Sepulveda}, A.~G. and {Wyatt}, M.~C. and {Matthews}, B.~C. and {MacGregor}, M.},
        title = "{REsolved ALMA and SMA Observations of Nearby Stars (REASONS): A population of 74 resolved planetesimal belts at millimetre wavelengths}",
      journal = {\aap},
         year = 2025,
        month = jan,
       volume = {693},
          eid = {A151},
        pages = {A151},
          doi = {10.1051/0004-6361/202451397},
archivePrefix = {arXiv},
       eprint = {2501.09058},
 primaryClass = {astro-ph.EP},
       adsurl = {https://ui.adsabs.harvard.edu/abs/2025A&A...693A.151M}
}

@ARTICLE{Iglesias2023winds,
       author = {{Iglesias}, Daniela P. and {Pani{\'c}}, Olja and {Rebollido}, Isabel},
        title = "{Gas absorption towards the {\ensuremath{\eta}} Tel debris disc: winds or clouds?}",
      journal = {\mnras},
         year = 2023,
        month = dec,
       volume = {526},
       number = {2},
        pages = {2500-2505},
          doi = {10.1093/mnras/stad2836},
archivePrefix = {arXiv},
       eprint = {2309.07746},
 primaryClass = {astro-ph.SR},
       adsurl = {https://ui.adsabs.harvard.edu/abs/2023MNRAS.526.2500I}
}

@ARTICLE{Somerville1988,
       author = {{Somerville}, W.~B.},
        title = "{The doublet-ratio method for interstellar abundances}",
      journal = {The Observatory},
         year = 1988,
        month = apr,
       volume = {108},
        pages = {44-49},
       adsurl = {https://ui.adsabs.harvard.edu/abs/1988Obs...108...44S}
}

@ARTICLE{Morton1991,
       author = {{Morton}, Donald C.},
        title = "{Atomic Data for Resonance Absorption Lines. I. Wavelengths Longward of the Lyman Limit}",
      journal = {\apjs},
         year = 1991,
        month = sep,
       volume = {77},
        pages = {119},
          doi = {10.1086/191601},
       adsurl = {https://ui.adsabs.harvard.edu/abs/1991ApJS...77..119M}
}

@INPROCEEDINGS{CastelliKurucz2003,
       author = {{Castelli}, F. and {Kurucz}, R.~L.},
        title = "{New Grids of ATLAS9 Model Atmospheres}",
    booktitle = {Modelling of Stellar Atmospheres},
         year = 2003,
       editor = {{Piskunov}, N. and {Weiss}, W.~W. and {Gray}, D.~F.},
       series = {IAU Symposium},
       volume = {210},
        month = jan,
        pages = {A20},
          doi = {10.48550/arXiv.astro-ph/0405087},
archivePrefix = {arXiv},
       eprint = {astro-ph/0405087},
 primaryClass = {astro-ph},
       adsurl = {https://ui.adsabs.harvard.edu/abs/2003IAUS..210P.A20C}
}

@ARTICLE{vizier, 
       author = {{Ochsenbein}, F. and {Bauer}, P. and {Marcout}, J.},
        title = "{The VizieR database of astronomical catalogues}",
      journal = {\aaps},
         year = 2000,
        month = apr,
       volume = {143},
        pages = {23-32},
          doi = {10.1051/aas:2000169},
archivePrefix = {arXiv},
       eprint = {astro-ph/0002122},
 primaryClass = {astro-ph},
       adsurl = {https://ui.adsabs.harvard.edu/abs/2000A&AS..143...23O}
}

@misc{Iglesias2020SearchingDisks,
  author       = {Iglesias, Daniela},
  title        = {Searching for gas in debris disks},
  month        = apr,
  year         = 2020,
  publisher    = {Zenodo},
  doi          = {10.5281/zenodo.14499893}
}

@article{Chen2003TheHerculis,
	author = {{Chen}, C.~H. and {Jura}, M.},
        title = "{The Low-Velocity Wind from the Circumstellar Matter around the B9 V Star {\ensuremath{\sigma}} Herculis}",
      journal = {\apj},
         year = 2003,
        month = jan,
       volume = {582},
       number = {1},
        pages = {443-448},
          doi = {10.1086/344589},
archivePrefix = {arXiv},
       eprint = {astro-ph/0209076},
 primaryClass = {astro-ph},
       adsurl = {https://ui.adsabs.harvard.edu/abs/2003ApJ...582..443C}
}

@article{Abt1973RotationDwarfs.,
	  author = {{Abt}, H.~A. and {Moyd}, K.~I.},
        title = "{Rotation and shell spectra among A-type dwarfs.}",
      journal = {\apj},
         year = 1973,
        month = jun,
       volume = {182},
        pages = {809},
          doi = {10.1086/152184},
       adsurl = {https://ui.adsabs.harvard.edu/abs/1973ApJ...182..809A}
}

@article{Welsh2013CircumstellarExocomets,
	author = {{Welsh}, Barry Y. and {Montgomery}, Sharon},
        title = "{Circumstellar Gas-Disk Variability Around A-Type Stars: The Detection of Exocomets?}",
      journal = {\pasp},
         year = 2013,
        month = jul,
       volume = {125},
       number = {929},
        pages = {759},
          doi = {10.1086/671757},
       adsurl = {https://ui.adsabs.harvard.edu/abs/2013PASP..125..759W}
}

@article{Koubsky1993ComingHerculis.,
	 author = {{Koubsky}, P. and {Horn}, J. and {Harmanec}, P. and {Hubert}, A.~M. and {Hubert}, H. and {Floquet}, M.},
        title = "{Coming shell phase of the Be star 4 Herculis.}",
      journal = {\aap},
         year = 1993,
        month = oct,
       volume = {277},
        pages = {521-523},
       adsurl = {https://ui.adsabs.harvard.edu/abs/1993A&A...277..521K}
}

@article{Welsh2018FurtherDiscs,
	author = {{Welsh}, Barry Y. and {Montgomery}, Sharon L.},
        title = "{Further detections of exocomet absorbing gas around Southern hemisphere A-type stars with known debris discs}",
      journal = {\mnras},
         year = 2018,
        month = feb,
       volume = {474},
       number = {2},
        pages = {1515-1525},
          doi = {10.1093/mnras/stx2800},
       adsurl = {https://ui.adsabs.harvard.edu/abs/2018MNRAS.474.1515W}
}

@article{Welsh1998Beta85905,
	author = {{Welsh}, B.~Y. and {Craig}, N. and {Crawford}, I.~A. and {Price}, R.~J.},
        title = "{Beta Pic-like circumstellar disk gas surrounding HR 10 and HD 85905}",
      journal = {\aap},
         year = 1998,
        month = oct,
       volume = {338},
        pages = {674-682},
       adsurl = {https://ui.adsabs.harvard.edu/abs/1998A&A...338..674W}
}

@ARTICLE{Haisch2001,
       author = {{Haisch}, Jr., Karl E. and {Lada}, Elizabeth A. and {Lada}, Charles J.},
        title = "{Disk Frequencies and Lifetimes in Young Clusters}",
      journal = {\apjl},
         year = 2001,
        month = jun,
       volume = {553},
       number = {2},
        pages = {L153-L156},
          doi = {10.1086/320685}
}

@ARTICLE{Hernandez2007,
       author = {{Hern{\'a}ndez}, Jes{\'u}s and {Hartmann}, L. and {Megeath}, T. and {Gutermuth}, R. and {Muzerolle}, J. and {Calvet}, N. and {Vivas}, A.~K. and {Brice{\~n}o}, C. and {Allen}, L. and {Stauffer}, J. and {Young}, E. and {Fazio}, G.},
        title = "{A Spitzer Space Telescope Study of Disks in the Young {\ensuremath{\sigma}} Orionis Cluster}",
      journal = {\apj},
         year = 2007,
        month = jun,
       volume = {662},
       number = {2},
        pages = {1067-1081},
          doi = {10.1086/513735},
}

@proceedings{PPVII2023,
  title        = {Protostars and Planets VII},
  editor       = {Inutsuka, Shu-ichiro and Aikawa, Yuri and Muto, Takayuki and Tomida, Kengo and Tamura, Motohide},
  booktitle    = {ASP Conference Series, Volume 534},
  year         = {2023},
  publisher    = {Astronomical Society of the Pacific},
  address      = {San Francisco, CA}
}

@article{Szewczyk2025,
    author = {Szewczyk, Karolina M and Panić, O and Iglesias, D P and Pearce, T D and Miley, J M},
    title = {HD 44892: The youngest (or oldest?) gas-harbouring debris disc around an intermediate mass star},
    journal = {Monthly Notices of the Royal Astronomical Society},
    pages = {staf2150},
    year = {2025},
    month = {12},
    issn = {0035-8711},
    doi = {10.1093/mnras/staf2150},
    url = {https://doi.org/10.1093/mnras/staf2150},
    eprint = {https://academic.oup.com/mnras/advance-article-pdf/doi/10.1093/mnras/staf2150/65727143/staf2150.pdf},
}

@INPROCEEDINGS{Lada1987,
       author = {{Lada}, Charles J.},
        title = "{Star formation: from OB associations to protostars.}",
    booktitle = {Star Forming Regions},
         year = 1987,
       editor = {{Peimbert}, Manuel and {Jugaku}, Jun},
       series = {IAU Symposium},
       volume = {115},
        month = jan,
        pages = {1},
}

@ARTICLE{Moor2025,
       author = {{Mo{\'o}r}, A. and {{\'A}brah{\'a}m}, P. and {K{\'o}sp{\'a}l}, {\'A}. and {Cataldi}, G. and {Hughes}, A.~M. and {Marino}, S. and {Kral}, Q. and {Milli}, J. and {Pawellek}, N.},
        title = "{Discovery of carbon monoxide emission from five debris disks around young A-type stars}",
      journal = {\aap},
         year = 2025,
        month = oct,
       volume = {703},
          eid = {A15},
        pages = {A15},
          doi = {10.1051/0004-6361/202554848}
}

@ARTICLE{Iglesias2025,
       author = {{Iglesias}, Daniela and {Rebollido}, Isabel and {Norazman}, Azib and {Snodgrass}, Colin and {Seligman}, Darryl Z. and {Xu}, Siyi and {Hoeijmakers}, H. Jens and {Kenworthy}, Matthew and {Lecavelier des Etangs}, Alain and {Bannister}, Michele and {Yang}, Bin},
        title = "{An Overview of Exocomets}",
      journal = {\ssr},
         year = 2025,
        month = dec,
       volume = {221},
       number = {8},
          eid = {122},
        pages = {122},
        doi = {10.1007/s11214-025-01247-6},
archivePrefix = {arXiv},
       eprint = {2511.08270},
 primaryClass = {astro-ph.EP},
       adsurl = {https://ui.adsabs.harvard.edu/abs/2025SSRv..221..122I}
}

@ARTICLE{Nakatani2023,
       author = {{Nakatani}, Riouhei and {Turner}, Neal J. and {Hasegawa}, Yasuhiro and {Cataldi}, Gianni and {Aikawa}, Yuri and {Marino}, Sebasti{\'a}n and {Kobayashi}, Hiroshi},
        title = "{A Primordial Origin for the Gas-rich Debris Disks around Intermediate-mass Stars}",
      journal = {\apjl},
         year = 2023,
        month = dec,
       volume = {959},
       number = {2},
          eid = {L28},
        pages = {L28},
        doi = {10.3847/2041-8213/ad0ed8},
archivePrefix = {arXiv},
       eprint = {2311.02195},
 primaryClass = {astro-ph.EP},
       adsurl = {https://ui.adsabs.harvard.edu/abs/2023ApJ...959L..28N}
}

@article{Rebollido2022The36546,
	author = {{Rebollido}, Isabel and {Ribas}, {\'A}lvaro and {de Gregorio-Monsalvo}, Itziar and {Villaver}, Eva and {Montesinos}, Benjam{\'\i}n and {Chen}, Christine and {Canovas}, H{\'e}ctor and {Henning}, Thomas and {Mo{\'o}r}, Attila and {Perrin}, Marshall and {Rivi{\`e}re-Marichalar}, Pablo and {Eiroa}, Carlos},
        title = "{The search for gas in debris discs: ALMA detection of CO gas in HD 36546}",
      journal = {\mnras},
         year = 2022,
        month = jan,
       volume = {509},
       number = {1},
        pages = {693-700},
          doi = {10.1093/mnras/stab2906},
archivePrefix = {arXiv},
       eprint = {2110.02308},
 primaryClass = {astro-ph.EP},
       adsurl = {https://ui.adsabs.harvard.edu/abs/2022MNRAS.509..693R}
}

@article{Matra2019On7,
	 author = {{Matr{\`a}}, L. and {{\"O}berg}, K.~I. and {Wilner}, D.~J. and {Olofsson}, J. and {Bayo}, A.},
        title = "{On the Ubiquity and Stellar Luminosity Dependence of Exocometary CO Gas: Detection around M Dwarf TWA 7}",
      journal = {\aj},
         year = 2019,
        month = mar,
       volume = {157},
       number = {3},
          eid = {117},
        pages = {117},
          doi = {10.3847/1538-3881/aaff5b},
archivePrefix = {arXiv},
       eprint = {1901.05004},
 primaryClass = {astro-ph.EP},
       adsurl = {https://ui.adsabs.harvard.edu/abs/2019AJ....157..117M}
}

@article{Marino2016ExocometaryRing,
	author = {{Marino}, S. and {Matr{\`a}}, L. and {Stark}, C. and {Wyatt}, M.~C. and {Casassus}, S. and {Kennedy}, G. and {Rodriguez}, D. and {Zuckerman}, B. and {Perez}, S. and {Dent}, W.~R.~F. and {Kuchner}, M. and {Hughes}, A.~M. and {Schneider}, G. and {Steele}, A. and {Roberge}, A. and {Donaldson}, J. and {Nesvold}, E.},
        title = "{Exocometary gas in the HD 181327 debris ring}",
      journal = {\mnras},
         year = 2016,
        month = aug,
       volume = {460},
       number = {3},
        pages = {2933-2944},
          doi = {10.1093/mnras/stw1216},
archivePrefix = {arXiv},
       eprint = {1605.05331},
 primaryClass = {astro-ph.EP},
       adsurl = {https://ui.adsabs.harvard.edu/abs/2016MNRAS.460.2933M}
}

@ARTICLE{Schneiderman2021,
       author = {{Schneiderman}, Tajana and {Matr{\`a}}, Luca and {Jackson}, Alan P. and {Kennedy}, Grant M. and {Kral}, Quentin and {Marino}, Sebasti{\'a}n and {{\"O}berg}, Karin I. and {Su}, Kate Y.~L. and {Wilner}, David J. and {Wyatt}, Mark C.},
        title = "{Carbon monoxide gas produced by a giant impact in the inner region of a young system}",
      journal = {\nat},
         year = 2021,
        month = oct,
       volume = {598},
       number = {7881},
        pages = {425-428},
          doi = {10.1038/s41586-021-03872-x},
archivePrefix = {arXiv},
       eprint = {2110.15377},
 primaryClass = {astro-ph.EP},
       adsurl = {https://ui.adsabs.harvard.edu/abs/2021Natur.598..425S}
}

@article{Riviere-Marichalar2014GasObservatory,
	author = {{Riviere-Marichalar}, P. and {Barrado}, D. and {Montesinos}, B. and {Duch{\^e}ne}, G. and {Bouy}, H. and {Pinte}, C. and {Menard}, F. and {Donaldson}, J. and {Eiroa}, C. and {Krivov}, A.~V. and {Kamp}, I. and {Mendigut{\'\i}a}, I. and {Dent}, W.~R.~F. and {Lillo-Box}, J.},
        title = "{Gas and dust in the beta Pictoris moving group as seen by the Herschel Space Observatory}",
      journal = {\aap},
         year = 2014,
        month = may,
       volume = {565},
          eid = {A68},
        pages = {A68},
          doi = {10.1051/0004-6361/201322901},
archivePrefix = {arXiv},
       eprint = {1404.1815},
 primaryClass = {astro-ph.SR},
       adsurl = {https://ui.adsabs.harvard.edu/abs/2014A&A...565A..68R}
}

@article{Moor2019New32297,
	author = {{Mo{\'o}r}, Attila and {Kral}, Quentin and {{\'A}brah{\'a}m}, P{\'e}ter and {K{\'o}sp{\'a}l}, {\'A}gnes and {Dutrey}, Anne and {Di Folco}, Emmanuel and {Hughes}, A. Meredith and {Juh{\'a}sz}, Attila and {Pascucci}, Ilaria and {Pawellek}, Nicole},
        title = "{New Millimeter CO Observations of the Gas-rich Debris Disks 49 Cet and HD 32297}",
      journal = {\apj},
         year = 2019,
        month = oct,
       volume = {884},
       number = {2},
          eid = {108},
        pages = {108},
          doi = {10.3847/1538-4357/ab4272},
archivePrefix = {arXiv},
       eprint = {1908.09685},
 primaryClass = {astro-ph.EP},
       adsurl = {https://ui.adsabs.harvard.edu/abs/2019ApJ...884..108M}
}

@article{Kospal2013ALMA21997,
	author = {{K{\'o}sp{\'a}l}, {\'A}. and {Mo{\'o}r}, A. and {Juh{\'a}sz}, A. and {{\'A}brah{\'a}m}, P. and {Apai}, D. and {Csengeri}, T. and {Grady}, C.~A. and {Henning}, Th. and {Hughes}, A.~M. and {Kiss}, Cs. and {Pascucci}, I. and {Schmalzl}, M.},
        title = "{ALMA Observations of the Molecular Gas in the Debris Disk of the 30 Myr Old Star HD 21997}",
      journal = {\apj},
         year = 2013,
        month = oct,
       volume = {776},
       number = {2},
          eid = {77},
        pages = {77},
          doi = {10.1088/0004-637X/776/2/77},
archivePrefix = {arXiv},
       eprint = {1310.5068},
 primaryClass = {astro-ph.SR},
       adsurl = {https://ui.adsabs.harvard.edu/abs/2013ApJ...776...77K}
}

@ARTICLE{Redfield2002,
       author = {{Redfield}, Seth and {Linsky}, Jeffrey L.},
        title = "{The Structure of the Local Interstellar Medium. I. High-Resolution Observations of Fe II, Mg II, and Ca II toward Stars within 100 Parsecs}",
      journal = {\apjs},
         year = 2002,
        month = apr,
       volume = {139},
       number = {2},
        pages = {439-465},
          doi = {10.1086/338650},
       adsurl = {https://ui.adsabs.harvard.edu/abs/2002ApJS..139..439R}
}

@ARTICLE{Olofsson2001,
       author = {{Olofsson}, G{\"o}ran and {Liseau}, Ren{\'e} and {Brandeker}, Alexis},
        title = "{Widespread Atomic Gas Emission Reveals the Rotation of the {\ensuremath{\beta}} Pictoris Disk}",
      journal = {\apjl},
         year = 2001,
        month = dec,
       volume = {563},
       number = {1},
        pages = {L77-L80},
          doi = {10.1086/338354},
archivePrefix = {arXiv},
       eprint = {astro-ph/0111206},
 primaryClass = {astro-ph},
       adsurl = {https://ui.adsabs.harvard.edu/abs/2001ApJ...563L..77O}
}

@ARTICLE{Roberge2000,
       author = {{Roberge}, A. and {Feldman}, P.~D. and {Lagrange}, A.~M. and {Vidal-Madjar}, A. and {Ferlet}, R. and {Jolly}, A. and {Lemaire}, J.~L. and {Rostas}, F.},
        title = "{High-Resolution Hubble Space Telescope STIS Spectra of C I and CO in the{\ensuremath{\beta}} Pictoris Circumstellar Disk}",
      journal = {\apj},
         year = 2000,
        month = aug,
       volume = {538},
       number = {2},
        pages = {904-910},
          doi = {10.1086/309157},
archivePrefix = {arXiv},
       eprint = {astro-ph/0003446},
 primaryClass = {astro-ph},
       adsurl = {https://ui.adsabs.harvard.edu/abs/2000ApJ...538..904R}
}

@ARTICLE{Brennan2024,
       author = {{Brennan}, Aoife and {Matr{\`a}}, Luca and {Marino}, Sebasti{\'a}n and {Wilner}, David and {Qi}, Chunhua and {Hughes}, A. Meredith and {Roberge}, Aki and {Hales}, Antonio S. and {Redfield}, Seth},
        title = "{Low CI/CO abundance ratio revealed by HST UV spectroscopy of CO-rich debris discs}",
      journal = {\mnras},
         year = 2024,
        month = jul,
       volume = {531},
       number = {4},
        pages = {4482-4502},
          doi = {10.1093/mnras/stae1328},
archivePrefix = {arXiv},
       eprint = {2405.13116},
 primaryClass = {astro-ph.EP},
       adsurl = {https://ui.adsabs.harvard.edu/abs/2024MNRAS.531.4482B}
}

@ARTICLE{Heays2017,
       author = {{Heays}, A.~N. and {Bosman}, A.~D. and {van Dishoeck}, E.~F.},
        title = "{Photodissociation and photoionisation of atoms and molecules of astrophysical interest}",
      journal = {\aap},
         year = 2017,
        month = jun,
       volume = {602},
          eid = {A105},
        pages = {A105},
          doi = {10.1051/0004-6361/201628742},
archivePrefix = {arXiv},
       eprint = {1701.04459},
 primaryClass = {astro-ph.SR},
       adsurl = {https://ui.adsabs.harvard.edu/abs/2017A&A...602A.105H}
}




\appendix

\section{Origin of narrow absorption features in full sample}

\begin{table*}
\caption{\label{tab:AllObjects}List of 130 objects from the \citet{Iglesias2023} sample used in this study, showing whether or not they have narrow absorption features in the Ca\,\textsc{ii} K line and how many, and if we consider them circumstellar or interstellar in nature. For parameters of these objects see the aforementioned study. Possible main-sequence stars are marked with $^{MS}$.}
\begin{tabular}{|l|l|l|l|l|l|}
\hline
Name & \# of absorption features & Abs feature origin & Name & \# of absorption features & Abs feature origin \\ \hline
HIP\,102719 & 1 & interstellar & HIP\,79878 & 1 & interstellar \\ \hline
HIP\,105148 & 1 & interstellar & HIP\,80387 & 1 & interstellar \\ \hline
HIP\,107962 & 1 & interstellar & HIP\,80477 & 0 & - \\ \hline
HIP\,115551 & 1 & interstellar & HIP\,81045 & 1 & interstellar \\ \hline
HIP\,12055 & 2 & interstellar & HIP\,81474 & 0 & - \\ \hline
HIP\,13910 & 1 & interstellar & HIP\,81891 & 2 & interstellar \\ \hline
HIP\,17543 & 1 & interstellar & HIP\,82091 & 0 & - \\ \hline
HIP\,21024 & 2 & interstellar & HIP\,82154 & 1 & interstellar \\ \hline
HIP\,22370 & 3 & interstellar & HIP\,82173 & 1 & interstellar \\ \hline
HIP\,22402 & 1 & interstellar &  HIP\,82714 & 1 & interstellar \\ \hline
HIP\,23633 & 1 & interstellar & HIP\,83505 & 2 & interstellar \\ \hline
HIP\,24092 & 1 & interstellar &  HIP\,84175 & 2 & interstellar \\ \hline
HIP\,25453 & 1 & interstellar & HIP\,84628 & 1 & interstellar \\ \hline
HIP\,25763 & 1 & interstellar &  HIP\,85372 & 0 & - \\ \hline
HIP\,25998 & 0 & - & HIP\,86429 & 1 & interstellar \\ \hline
HIP\,26783 & 1 & interstellar & HIP\,87807 & 2 & interstellar \\ \hline
HIP\,28877 & 3 & interstellar & HIP\,89114 & 0 & - \\ \hline
HIP\,30054 & 4 & interstellar & HIP\,90609 & 3 & interstellar \\ \hline
HIP\,30353 & 2 & interstellar & HIP\,91779 & 1 & interstellar \\ \hline
HIP\,30414 & variable features & possibly circumstellar &  HIP\,91893 & 1 & interstellar \\ \hline
HIP\,30672 & 3 & interstellar & HIP\,92188 & 1 & interstellar \\ \hline
HIP\,32667 & 2 & interstellar & HIP\,92800 & 1 & interstellar \\ \hline
HIP\,32731 & 2 & interstellar &  HIP\,93525 & 1 & interstellar \\ \hline
HIP\,36225 & 1 & interstellar & HIP\,93542 & 1 & interstellar \\ \hline
HIP\,36966 & 1 & interstellar &  HIP\,93744 & 0 & -  \\ \hline
HIP\,37094 & 1 & interstellar & HIP\,94167 & 2 & interstellar \\ \hline
HIP\,37697 & 1 & interstellar & HIP\,94949 & 2 & interstellar \\ \hline
HIP\,40594 & 1 & interstellar & HIP\,95547 & 1 & interstellar \\ \hline
HIP\,43643 & 1 & interstellar & HIP\,95619 & 2 & interstellar \\ \hline
HIP\,45585 & 1 & interstellar & HIP\,95979 & 2 & interstellar \\ \hline
HIP\,48613 & 3 & interstellar &  TYC\,1019-473-1$^{MS}$ & 0 & - \\ \hline
HIP\,49606 & 1 & interstellar &  TYC\,1302-69-1 & 3 & interstellar \\ \hline
HIP\,51843 & 1 & interstellar & TYC\,1600-168-1 & 2 & interstellar \\ \hline
HIP\,52688 & 1 & interstellar &   TYC\,1765-1373-1 & 1 & interstellar\\ \hline
HIP\,56943 & 2 & interstellar &  TYC\,2138-1154-1 & 0 & - \\ \hline
HIP\,57027 & 0 & - & TYC\,2140-893-1$^{MS}$ & 2 & interstellar \\ \hline
HIP\,58720 & 1 & interstellar & TYC\,435-302-1 & 2 & interstellar \\ \hline
HIP\,59898 & 1 & interstellar & TYC\,457-39-1 & 1 & interstellar \\ \hline
HIP\,60183 & 1 & interstellar & TYC\,4822-542-1 & 1 & interstellar \\ \hline
HIP\,60561 & 1 & interstellar & TYC\,5124-2609-1 & 2 & interstellar \\ \hline
HIP\,61498 & 2 & interstellar & TYC\,5597-720-1 & 1 & interstellar \\ \hline
HIP\,61782$^{MS}$ & 2 & circumstellar &  TYC\,5649-822-1 & 3 & interstellar \\ \hline
HIP\,63839 & 0 & - & TYC\,6487-537-1 & 1 & interstellar \\ \hline
HIP\,66068 & 1 & interstellar & TYC\,6822-283-1 & 5 & inconclusive \\ \hline
HIP\,69011 & 1 & interstellar & TYC\,6843-956-1 & 1 & interstellar \\ \hline
HIP\,69761 & 1 & interstellar & TYC\,6846-840-1 & 1 & interstellar \\ \hline
HIP\,70040 & 1 & interstellar & TYC\,6849-1024-1 & 1 & interstellar \\ \hline
HIP\,73150 & 1 & interstellar & TYC\,6849-1059-1 & 1 & interstellar \\ \hline
HIP\,73397$^{MS}$ & 1 & interstellar & TYC\,7374-702-1 & 1 & interstellar \\ \hline
HIP\,75210 & 2 & interstellar & TYC\,7379-170-1 & 1 & interstellar \\ \hline
HIP\,75902 & 1 & interstellar & TYC\,7389-385-1 & 1 & interstellar \\ \hline
HIP\,76011 & 1 & interstellar &  TYC\,7645-1683-1 & 2 & interstellar \\ \hline
HIP\,76234 & 4 & interstellar & TYC\,7851-91-1 & 2 & interstellar \\ \hline
HIP\,76244 & 1 & interstellar & TYC\,7879-1373-1 & 3 & possibly circumstellar  \\ \hline
HIP\,76310 & 1 & interstellar & TYC\,7879-1506-1 & 1 & interstellar \\ \hline
HIP\,76395 & 1 & interstellar  & TYC\,7892-4279-1 & 0 & -  \\ \hline
HIP\,77315 & 0 & - & TYC\,8161-658-1 & 1 & interstellar\\ \hline
HIP\,77317 & 0 & - & TYC\,8241-3236-1 & 1 & interstellar \\ \hline
HIP\,77562 & 1 & interstellar &  TYC\,8603-1782-1 & 1 & interstellar \\ \hline
HIP\,77911 & 1 & interstellar & TYC\,8697-1251-1 & 0 & -  \\ \hline
HIP\,79179 & 1 & interstellar & TYC\,8703-571-1 & 1 & interstellar \\ \hline
\end{tabular}
\end{table*}

\begin{table}
\begin{tabular}{|l|l|l|}
\hline
Name & \# of absorption features & Abs feature origin \\ \hline
TYC\,8946-1580-1 & 1 & interstellar \\ \hline
TYC\,8959-774-1 & 1 & interstellar \\ \hline
TYC\,8977-9252-1 & 1 & interstellar \\ \hline
TYC\,8995-1696-1 & 1 & interstellar \\ \hline
TYC\,9004-2647-1 & 1 & interstellar \\ \hline
TYC\,9005-873-1 & 1 & interstellar \\ \hline
TYC\,9009-1874-1 & 1 & interstellar \\ \hline
TYC\,9329-60-1 & 1 & interstellar \\ \hline
\end{tabular}
\end{table}

\FloatBarrier

\section{Observations}

\begin{onecolumn}
\begin{longtable}{|l|l|l|l|l|l|l|l|}
\captionsetup{width=\textwidth}
\caption{\label{tab:ObservationDetails} Observation details for the 130 objects analysed in this study. The columns include: object name used in this study, HD identifier (if available), coordinates, instruments used for observations, number of observations, observation dates, and signal-to-noise ratio (maximum value if multiple epochs are available).} \\
\hline
Name & HD identifier & RA & Dec & Instrument & Number of spectra & Observation dates & S/N \\ \hline
\endfirsthead

\hline
Name & HD identifier & RA & Dec & Instrument & Number of spectra & Observation dates & S/N \\ \hline
\endhead

\multicolumn{8}{r}{\textit{Continued on next page}} \\ 
\endfoot
\endlastfoot
    
HIP\,102719 & HD\,197827 & 20:48:50.68 & -59:14:04.0 & UVES & 6 & 2022-07-10 to 2022-08-22 & 160 \\ \hline
HIP\,105148 & HD\,202299 & 21:18:00.49 & -64:40:53.7 & FEROS & 1 & 2007-05-25 to 2007-05-25 & 359 \\ \hline
HIP\,107962 & HD\,207888 & 21:52:21.25 & -03:10:28.8 & UVES & 4 & 2022-05-11 to 2022-07-05 & 191 \\ \hline
HIP\,115551 & HD\,220529 & 23:24:21.80 & +06:11:10.6 & UVES & 4 & 2022-07-05 to 2022-08-23 & 170 \\ \hline
HIP\,12055 & HD\,16152 & 02:35:24.47 & -09:21:02.7 & UVES & 4 & 2022-07-05 to 2022-08-23 & 129 \\ \hline
HIP\,13910 & HD\,18572 & 02:59:08.41 & -04:47:00.1 & UVES & 4 & 2022-09-01 to 2022-09-11 & 123 \\ \hline
HIP\,17543 & HD\,24188 & 03:45:23.74 & -71:39:29.3 & UVES & 6 & 2023-09-13 to 2023-09-15 & 323 \\ \hline
HIP\,21024 & HD\,28813 & 04:30:29.54 & -43:24:38.2 & UVES & 8 & 2023-09-13 to 2024-02-02 & 274 \\ \hline
HIP\,22370 & HD\,30861 & 04:48:57.47 & -47:08:04.3 & UVES & 6 & 2022-07-24 to 2022-08-23 & 139 \\ \hline
HIP\,22402 & HD\,30466 & 04:49:16.01 & +29:34:16.8 & UVES & 6 & 2023-09-20 to 2023-09-23 & 249 \\ \hline
HIP\,23633 & HD\,32509 & 05:04:50.13 & +26:43:14.8 & UVES & 6 & 2023-09-20 to 2023-10-11 & 238 \\ \hline
HIP\,24092 & HD\,33594 & 05:10:32.96 & -18:39:44.8 & UVES & 4 & 2022-09-01 to 2022-09-09 & 116 \\ \hline
HIP\,25453 & HD\,35656 & 05:26:38.83 & +06:52:07.2 & UVES & 2 & 2022-09-23 to 2022-09-23 & 157 \\ \hline
HIP\,25763 & HD\,36044C & 05:30:05.99 & +29:33:10.1 & UVES & 12 & 2023-10-11 to 2023-10-16 & 356 \\ \hline
HIP\,25998 & HD\,37004 & 05:32:56.39 & -47:41:20.9 & UVES & 6 & 2022-07-24 to 2022-09-09 & 127 \\ \hline
HIP\,26783 & HD\,37647 & 05:41:21.47 & +29:29:39.9 & UVES & 6 & 2023-10-11 to 2023-10-12 & 314 \\ \hline
HIP\,28877 & HD\,41649 & 06:05:48.33 & -16:57:33.8 & UVES & 6 & 2023-09-13 to 2023-09-27 & 206 \\ \hline
HIP\,30054 & HD\,44405 & 06:19:28.47 & -42:53:10.9 & UVES & 6 & 2023-09-13 to 2023-09-17 & 201 \\ \hline
HIP\,30353 & HD\,44618 & 06:23:05.33 & +10:10:03.7 & UVES & 6 & 2023-09-21 to 2023-09-29 & 241 \\ \hline
HIP\,30414 & HD\,44892 & 06:23:42.76 & -16:28:01.9 & UVES & 6 & 2023-09-13 to 2023-09-27 & 291 \\ \hline
HIP\,30672 & HD\,45681 & 06:26:43.63 & -44:04:02.9 & UVES & 4 & 2022-10-20 to 2022-12-06 & 213 \\ \hline
HIP\,32667 & HD\,49610 & 06:48:49.30 & -08:18:36.9 & UVES & 6 & 2023-10-16 to 2023-11-01 & 255 \\ \hline
HIP\,32731 & HD\,49812 & 06:49:40.11 & -15:02:40.5 & UVES & 5 & 2022-12-29 to 2023-10-16 & 246 \\ \hline
HIP\,36225 & HD\,59074 & 07:27:34.48 & -18:29:30.8 & UVES & 6 & 2023-10-16 to 2023-11-01 & 266 \\ \hline
HIP\,36966 & HD\,60875 & 07:35:56.95 & -18:50:47.0 & UVES & 6 & 2023-10-16 to 2023-11-11 & 391 \\ \hline
HIP\,37094 & HD\,61308 & 07:37:20.03 & -30:47:32.0 & UVES & 9 & 2023-10-16 to 2023-10-30 & 259 \\ \hline
HIP\,37697 & HD\,62803 & 07:44:02.50 & -38:49:04.6 & UVES & 6 & 2022-09-08 to 2022-09-19 & 149 \\ \hline
HIP\,40594 & HD\,69952 & 08:17:16.65 & -46:53:35.7 & UVES & 2 & 2022-09-14 to 2022-09-14 & 138 \\ \hline
HIP\,43643 & HD\,76268 & 08:53:22.31 & -44:19:59.2 & UVES & 2 & 2022-09-25 to 2022-09-25 & 176 \\ \hline
HIP\,45585 & HD\,80950 & 09:17:27.57 & -74:44:04.5 & HARPS & 28 & 2012-02-19 to 2012-05-03 & 175 \\ \hline
HIP\,48613 & HD\,86087 & 09:54:51.23 & -50:14:38.3 & FEROS & 2 & 2016-01-03 to 2016-01-04 & 306 \\ 
 &  &  &  & HARPS & 21 & 2018-12-19 to 2019-07-19 & 385 \\ 
 &  &  &  & UVES & 2 & 2019-01-24 to 2019-01-24 & 232 \\ \hline
HIP\,49606 & HD\,88068 & 10:07:31.00 & -57:32:59.8 & UVES & 6 & 2023-10-16 to 2023-11-13 & 423 \\ \hline
HIP\,51843 & HD\,92047 & 10:35:30.32 & -69:49:02.8 & UVES & 4 & 2022-06-11 to 2022-07-09 & 127 \\ \hline
HIP\,52688 & HD\,93331 & 10:46:24.83 & -13:27:35.5 & UVES & 6 & 2023-11-22 to 2023-11-30 & 267 \\ \hline
HIP\,56943 & HD\,101511 & 11:40:26.42 & -62:11:53.8 & UVES & 2 & 2022-07-10 to 2022-07-10 & 153 \\ \hline
HIP\,57027 & HD\,101727 & 11:41:32.57 & -77:03:16.9 & UVES & 2 & 2022-07-10 to 2022-07-10 & 199 \\ \hline
HIP\,58720 & HD\,104600 & 12:02:37.69 & -69:11:32.2 & UVES & 6 & 2017-02-22 to 2019-02-21 & 239 \\ \hline
HIP\,59898 & HD\,106797 & 12:17:06.31 & -65:41:34.7 & UVES & 6 & 2017-02-22 to 2019-02-21 & 289 \\ \hline
HIP\,60183 & HD\,107301 & 12:20:28.22 & -65:50:33.6 & UVES & 6 & 2017-02-22 to 2019-02-21 & 261 \\ \hline
HIP\,60561 & HD\,107947 & 12:24:51.92 & -72:36:14.0 & UVES & 8 & 2018-12-24 to 2018-12-28 & 284 \\ \hline
HIP\,61498 & HD\,109573 & 12:36:01.03 & -39:52:10.2 & FEROS & 29 & 2007-03-09 to 2017-04-09 & 492 \\ 
 &  & &  & HARPS & 10 & 2007-12-06 to 2015-01-21 & 313 \\ 
 &  & &  & UVES & 486 & 2002-01-19 to 2007-05-15 & 523 \\ \hline
HIP\,61782 & HD\,110058 & 12:39:46.20 & -49:11:55.6 & FEROS & 11 & 2010-01-31 to 2017-04-09 & 329 \\ 
 &  & &  & HARPS & 15 & 2015-01-18 to 2018-03-31 & 205 \\ \hline
HIP\,63839 & HD\,113457 & 13:05:02.04 & -64:26:29.7 & UVES & 6 & 2017-03-20 to 2019-02-20 & 358 \\ \hline
HIP\,66068 & HD\,117665 & 13:32:39.24 & -44:27:00.9 & UVES & 6 & 2019-01-24 to 2019-02-21 & 251 \\ \hline
HIP\,69011 & HD\,123247 & 14:07:40.81 & -48:42:14.5 & UVES & 6 & 2023-12-20 to 2023-12-27 & 208 \\ \hline
HIP\,69761 & HD\,124564 & 14:16:37.50 & -62:06:59.4 & UVES & 4 & 2022-05-26 to 2022-07-10 & 221 \\ \hline
HIP\,70040 & HD\,125205 & 14:19:54.59 & -57:09:27.3 & UVES & 6 & 2023-12-28 to 2023-12-31 & 181 \\ \hline
HIP\,73150 & HD\,131885 & 14:56:57.45 & -26:17:06.1 & UVES & 4 & 2022-05-26 to 2022-07-13 & 189 \\ \hline
HIP\,73397 & HD\,132658 & 14:59:59.53 & +12:01:26.2 & UVES & 4 & 2022-08-05 to 2022-08-22 & 163 \\ \hline
HIP\,75210 & HD\,136482 & 15:22:11.25 & -37:38:08.2 & HARPS & 1 & 2015-01-19 to 2015-01-19 & 192 \\ 
 &  &  & & UVES & 4 & 2017-03-20 to 2019-02-24 & 322 \\ \hline
HIP\,75902 & HD\,137641 & 15:30:11.59 & -60:29:27.2 & UVES & 4 & 2022-07-10 to 2022-08-10 & 138 \\ \hline
HIP\,76011 & HD\,137509 & 15:31:27.11 & -71:03:43.7 & UVES & 6 & 2001-07-12 to 2008-04-17 & 175 \\ \hline
HIP\,76234 & HD\,138564 & 15:34:20.85 & -39:20:57.1 & UVES & 6 & 2023-09-26 to 2024-01-11 & 215 \\ \hline
HIP\,76244 & HD\,138334 & 15:34:26.59 & -59:48:03.4 & UVES & 4 & 2022-06-11 to 2022-07-10 & 137 \\ \hline
HIP\,76310 & HD\,138813 & 15:35:16.11 & -25:44:03.0 & FEROS & 13 & 2016-03-26 to 2017-04-08 & 389 \\ 
 &  & &  & UVES & 6 & 2016-02-22 to 2016-03-18 & 181 \\ \hline
HIP\,76395 & HD\,138923 & 15:36:11.36 & -33:05:34.1 & FEROS & 2 & 2009-02-06 to 2009-02-06 & 233 \\ 
 &  &  &  & UVES & 6 & 2019-01-26 to 2019-02-22 & 253 \\ \hline
HIP\,77315 & HD\,140817 & 15:47:04.46 & -35:30:37.2 & FEROS & 2 & 2009-07-20 to 2009-07-20 & 210 \\ 
 &  &  &  & UVES & 6 & 2019-02-24 to 2019-03-10 & 336 \\ \hline
HIP\,77317 & HD\,140840 & 15:47:06.17 & -35:31:04.9 & FEROS & 2 & 2009-02-06 to 2009-02-06 & 213 \\ 
 &  &  &  & UVES & 1 & 2016-02-21 to 2016-02-21 & 126 \\ \hline
HIP\,77562 & HD\,141168 & 15:50:07.08 & -53:12:35.2 & UVES & 4 & 2022-07-10 to 2022-07-13 & 187 \\ \hline
HIP\,77911 & HD\,142315 & 15:54:41.60 & -22:45:58.5 & FEROS & 5 & 2010-02-01 to 2017-04-05 & 333 \\ \hline
HIP\,79179 & HD\,144696 & 16:09:38.67 & -52:06:55.1 & UVES & 4 & 2022-07-13 to 2022-07-23 & 186 \\ \hline
HIP\,79878 & HD\,146606 & 16:18:16.16 & -28:02:30.1 & FEROS & 5 & 2009-07-20 to 2017-04-09 & 309 \\ 
 &  &  &  & UVES & 2 & 2016-09-12 to 2016-09-12 & 220 \\ \hline
HIP\,80387 & HD\,147703 & 16:24:30.43 & -27:09:04.2 & UVES & 4 & 2022-07-13 to 2022-07-23 & 162 \\ \hline
HIP\,80477 & HD\,148021 & 16:25:38.21 & -10:05:02.8 & UVES & 6 & 2023-09-23 to 2024-01-31 & 213 \\ \hline
HIP\,81045 & HD\,148850 & 16:33:11.00 & -47:41:00.2 & UVES & 4 & 2022-07-13 to 2022-07-23 & 214 \\ \hline
HIP\,81474 & HD\,149914 & 16:38:28.65 & -18:13:13.7 & UVES & 6 & 2023-09-17 to 2023-09-30 & 278 \\ \hline
HIP\,81891 & HD\,150638 & 16:43:38.72 & -32:06:21.4 & UVES & 6 & 2023-09-14 to 2023-09-17 & 217 \\ \hline
HIP\,82091 & HD\,150897 & 16:46:06.37 & -46:31:55.2 & UVES & 4 & 2022-07-13 to 2022-07-25 & 181 \\ \hline
HIP\,82154 & HD\,151109 & 16:47:01.68 & -39:32:01.9 & FEROS & 2 & 2007-04-01 to 2017-07-19 & 359 \\ 
 & & & & UVES & 4 & 2016-10-01 to 2016-10-01 & 287 \\ \hline
HIP\,82173 & HD\,150915 & 16:47:19.90 & -57:27:38.9 & UVES & 5 & 2022-07-25 to 2022-08-14 & 181 \\ \hline
HIP\,82714 & HD\,152384 & 16:54:26.39 & -33:28:31.3 & UVES & 6 & 2022-07-14 to 2022-07-25 & 162 \\ \hline
HIP\,83505 & HD\,154002 & 17:03:59.08 & -28:15:22.7 & FEROS & 1 & 2008-04-01 to 2008-04-01 & 233 \\
 &  & & & UVES & 4 & 2022-06-24 to 2022-07-14 & 115 \\ \hline
HIP\,84175 & HD\,155401 & 17:12:25.06 & -27:45:43.7 & UVES & 6 & 2022-07-14 to 2022-07-25 & 189 \\ \hline
HIP\,84628 & HD\,156157 & 17:18:01.37 & -43:34:39.7 & UVES & 8 & 2022-07-09 to 2022-07-23 & 196 \\ \hline
HIP\,85372 & HD\,157751 & 17:26:41.19 & -34:05:32.5 & HARPS & 3 & 2007-04-15 to 2007-04-15 & 171 \\ \hline
HIP\,86429 & HD\,160047 & 17:39:34.62 & -38:23:44.5 & UVES & 4 & 2023-03-11 to 2023-03-23 & 240 \\ \hline
HIP\,87807 & HD\,163422 & 17:56:15.03 & -04:34:35.0 & UVES & 6 & 2023-09-14 to 2023-09-15 & 276 \\ \hline
HIP\,89114 & HD\,166393 & 18:11:14.78 & -19:50:31.0 & UVES & 6 & 2022-07-17 to 2022-08-11 & 156 \\ \hline
HIP\,90609 & HD\,170562 & 18:29:20.86 & +17:59:02.1 & UVES & 6 & 2023-09-14 to 2023-09-16 & 196 \\ \hline
HIP\,91779 & HD\,173054 & 18:42:54.22 & +13:34:00.2 & UVES & 4 & 2022-08-10 to 2022-08-11 & 141 \\ \hline
HIP\,91893 & HD\,173196 & 18:43:59.29 & +03:20:38.2 & UVES & 4 & 2022-08-06 to 2022-08-10 & 135 \\ \hline
HIP\,92188 & HD\,173505 & 18:47:19.04 & -37:32:55.2 & FEROS & 1 & 2019-08-01 to 2019-08-01 & 90 \\ 
 &  &  &  & UVES & 4 & 2014-04-27 to 2014-04-27 & 184 \\ \hline
HIP\,92800 & HD\,175427 & 18:54:32.85 & +20:36:55.1 & UVES & 6 & 2023-09-14 to 2023-09-16 & 188 \\ \hline
HIP\,93525 & HD\,177064 & 19:02:53.62 & +05:20:33.9 & UVES & 4 & 2022-08-19 to 2022-08-21 & 151 \\ \hline
HIP\,93542 & HD\,176638 & 19:03:06.88 & -42:05:42.4 & FEROS & 2 & 2009-06-02 to 2010-05-26 & 405 \\
 &  & &  & HARPS & 10 & 2005-08-20 to 2006-05-25 & 273 \\ \hline
HIP\,93744 & HD\,177625 & 19:05:20.03 & +05:53:37.6 & UVES & 4 & 2022-08-06 to 2022-08-10 & 191 \\ \hline
HIP\,94167 & HD\,178857 & 19:10:07.28 & +00:55:55.2 & UVES & 4 & 2022-08-06 to 2022-08-14 & 170 \\ \hline
HIP\,94949 & HD\,181253 & 19:19:17.01 & +14:10:39.3 & UVES & 6 & 2023-09-14 to 2023-09-16 & 321 \\ \hline
HIP\,95547 & HD\,182740 & 19:26:05.13 & +03:31:09.7 & UVES & 4 & 2022-08-11 to 2022-08-14 & 152 \\ \hline
HIP\,95619 & HD\,182681 & 19:26:56.48 & -29:44:35.6 & UVES & 6 & 2023-09-13 to 2023-09-14 & 257 \\ \hline
HIP\,95979 & HD\,182254 & 19:31:01.04 & -73:59:06.8 & UVES & 6 & 2022-07-09 to 2022-07-13 & 183 \\ \hline
TYC\,1019-473-1 & HD\,162951 & 17:53:08.80 & +14:00:34.9 & UVES & 4 & 2022-08-11 to 2022-08-26 & 232 \\ \hline
TYC\,1302-69-1 & HD\,37673 & 05:41:08.87 & +18:37:28.0 & UVES & 6 & 2023-10-12 to 2023-10-16 & 258 \\ \hline
TYC\,1600-168-1 & HD\,182445 & 19:24:03.98 & +15:43:46.5 & UVES & 4 & 2022-08-06 to 2022-08-10 & 123 \\ \hline
TYC\,1765-1373-1 & - & 02:21:36.14 & +23:37:54.7 & UVES & 12 & 2023-09-13 to 2023-09-14 & 232 \\ \hline
TYC\,2138-1154-1 & HD\,344687 & 19:39:26.96 & +23:52:48.0 & UVES & 4 & 2022-08-10 to 2022-08-19 & 189 \\ \hline
TYC\,2140-893-1 & HD\,345089 & 19:52:03.18 & +23:26:21.1 & UVES & 4 & 2022-08-10 to 2022-08-13 & 159 \\ \hline
TYC\,365-704-1 & HD\,142537 & 15:56:36.40 & -42:19:33.8 & FEROS & 1 & 2008-06-16 to 2008-06-16 & 395 \\ \hline
TYC\,435-302-1 & HD\,167392 & 18:14:52.60 & +03:40:23.3 & UVES & 4 & 2022-08-10 to 2022-08-11 & 141 \\ \hline
TYC\,457-39-1 & HD\,176305 & 18:59:25.67 & +05:15:03.5 & UVES & 4 & 2022-08-10 to 2022-08-22 & 143 \\ \hline
TYC\,4703-768-1 & HD\,18572 & 02:59:08.41 & -04:47:00.1 & UVES & 4 & 2022-09-01 to 2022-09-11 & 123 \\ \hline
TYC\,4822-542-1 & HD\,52208 & 07:00:09.45 & -05:22:41.8 & UVES & 6 & 2023-10-02 to 2023-10-16 & 243 \\ \hline
TYC\,5091-298-1 & HD\,163422 & 17:56:15.03 & -04:34:35.0 & UVES & 6 & 2023-09-14 to 2023-09-15 & 276 \\ \hline
TYC\,5124-2609-1 & HD\,171837 & 18:37:14.68 & -06:36:17.8 & UVES & 12 & 2023-09-13 to 2023-09-15 & 315 \\ \hline
TYC\,5597-720-1 & HD\,140541 & 15:44:30.83 & -09:13:55.0 & UVES & 4 & 2023-02-28 to 2023-03-20 & 246 \\ \hline
TYC\,5649-822-1 & HD\,156267 & 17:17:02.17 & -10:39:42.3 & UVES & 4 & 2022-08-06 to 2022-08-09 & 129 \\ \hline
TYC\,5905-1186-1 & HD\,33594 & 05:10:32.96 & -18:39:44.8 & UVES & 4 & 2022-09-01 to 2022-09-09 & 116 \\ \hline
TYC\,6487-537-1 & - & 05:20:35.61 & -29:43:29.0 & UVES & 12 & 2023-09-13 to 2023-09-17 & 250 \\ \hline
TYC\,6822-283-1 & HD\,152519 & 16:55:00.67 & -29:51:10.2 & UVES & 6 & 2022-07-14 to 2022-07-25 & 165 \\ \hline
TYC\,6843-956-1 & HD\,165999 & 18:09:36.27 & -23:34:04.7 & UVES & 5 & 2023-03-11 to 2023-09-14 & 233 \\ \hline
TYC\,6846-840-1 & HD\,164147 & 18:00:50.96 & -24:47:18.7 & UVES & 6 & 2022-06-24 to 2022-08-26 & 190 \\ \hline
TYC\,6849-1024-1 & HD\,314750 & 17:53:04.53 & -27:00:12.7 & UVES & 4 & 2022-06-24 to 2022-07-23 & 162 \\ \hline
TYC\,6849-1059-1 & HD\,162655 & 17:53:16.77 & -27:37:44.3 & UVES & 4 & 2022-06-24 to 2022-07-10 & 124 \\ \hline
TYC\,7374-702-1 & HD\,157036 & 17:22:31.48 & -36:30:49.6 & UVES & 6 & 2022-07-14 to 2022-07-25 & 153 \\ \hline
TYC\,7379-170-1 & HD\,158422 & 17:30:40.21 & -32:03:42.3 & UVES & 4 & 2022-06-24 to 2022-07-14 & 110 \\ \hline
TYC\,7389-385-1 & HD\,161759 & 17:48:59.86 & -37:17:42.8 & UVES & 4 & 2022-06-24 to 2022-07-09 & 151 \\ \hline
TYC\,7645-1683-1 & HD\,62938 & 07:44:44.89 & -38:03:13.4 & UVES & 4 & 2022-09-14 to 2022-09-20 & 147 \\ \hline
TYC\,7851-91-1 & HD\,143181 & 16:00:22.34 & -38:59:37.7 & UVES & 4 & 2022-07-13 to 2022-07-23 & 160 \\ \hline
TYC\,7879-1373-1 & HD\,151183 & 16:47:47.43 & -43:56:18.1 & UVES & 4 & 2022-07-13 to 2022-07-22 & 148 \\ \hline
TYC\,7879-1506-1 & HD\,150477 & 16:43:11.25 & -43:39:10.4 & UVES & 4 & 2022-07-13 to 2022-07-23 & 166 \\ \hline
TYC\,7892-4279-1 & HD\,159807 & 17:38:36.75 & -42:44:01.9 & UVES & 4 & 2022-06-24 to 2022-07-10 & 154 \\ \hline
TYC\,8161-658-1 & HD\,70702 & 08:21:00.82 & -51:32:36.9 & FEROS & 1 & 2007-02-10 to 2007-02-10 & 123 \\ \hline
TYC\,8241-3236-1 & HD\,105383 & 12:08:04.87 & -50:45:49.1 & UVES & 2 & 2022-07-10 to 2022-07-10 & 144 \\ \hline
TYC\,8603-1782-1 & HD\,87322 & 10:02:31.48 & -56:12:22.9 & UVES & 2 & 2022-07-09 to 2022-07-09 & 143 \\ \hline
TYC\,8697-1251-1 & HD\,142564 & 15:57:51.79 & -53:45:27.1 & UVES & 4 & 2022-07-10 to 2022-07-13 & 143 \\ \hline
TYC\,8703-571-1 & HD\,136155 & 15:21:34.13 & -56:54:07.3 & UVES & 4 & 2022-07-10 to 2022-07-23 & 171 \\ \hline
TYC\,8946-1580-1 & HD\,86836 & 09:58:57.89 & -62:19:00.2 & UVES & 2 & 2022-07-09 to 2022-07-09 & 156 \\ \hline
TYC\,8959-774-1 & HD\,97558 & 11:12:38.20 & -60:55:16.1 & UVES & 2 & 2022-07-10 to 2022-07-10 & 105 \\ \hline
TYC\,8977-9252-1 & HD\,103457 & 11:54:32.75 & -63:42:02.2 & UVES & 2 & 2022-07-10 to 2022-07-10 & 155 \\ \hline
TYC\,8995-1696-1 & HD\,117959 & 13:35:26.97 & -61:57:09.3 & UVES & 4 & 2022-05-30 to 2022-07-09 & 121 \\ \hline
TYC\,9004-2647-1 & HD\,121808 & 13:59:40.59 & -61:22:52.5 & UVES & 4 & 2022-05-30 to 2022-07-09 & 129 \\ \hline
TYC\,9005-873-1 & HD\,124340 & 14:15:09.92 & -61:06:42.1 & UVES & 4 & 2022-07-09 to 2022-08-10 & 127 \\ \hline
TYC\,9009-1874-1 & HD\,122965 & 14:07:05.94 & -62:16:38.0 & UVES & 4 & 2022-05-26 to 2022-05-30 & 190 \\ \hline
TYC\,9329-60-1 & HD\,198147 & 20:53:24.09 & -71:50:13.9 & UVES & 12 & 2023-09-13 to 2023-09-14 & 241 \\ \hline
\end{longtable}

\pagebreak

\begin{longtable}{|l|l|l|l|l|l|l|}
\captionsetup{width=0.85\textwidth}
\caption{\label{tab:ObservationDetailsNearby} Observation details for the additional 38 objects used as comparison stars in this study. The columns include: object name used in this study, coordinates, instruments used for observations, number of observations, observation dates, and signal-to-noise ratio (maximum value if multiple epochs are available).} \\
\hline
Name & RA & Dec & Instrument & Number of spectra & Observation dates & S/N \\ \hline
\endfirsthead
BD-07\,534 & 12:20:11.44 & -53:05:22.4 & UVES & 2 & 2024-08-12 & 364 \\ \hline
HD\,101638 & 19:49:3.71 & -09:04:22.6 & UVES & 2 & 2024-05-22 & 302 \\ \hline
HD\,101671 & 19:40:5.71 & +00:58:14.8 & UVES & 2 & 2024-05-21 & 226 \\ \hline
HD\,117337 & 20:41:13.50 & +18:26:18.9 & UVES & 2 & 2024-05-20 & 368 \\ \hline
HD\,121720 & 20:43:31.22 & +00:12:52.3 & UVES & 2 & 2024-05-22 & 356 \\ \hline
HD\,131369 & 22:16:36.73 & +28:36:33.9 & UVES & 2 & 2024-06-22 & 150 \\ \hline
HD\,132446 & 00:49:55.89 & +56:34:24.1 & UVES & 2 & 2024-08-17 & 374 \\ \hline
HD\,136423 & 22:12:49.75 & +16:19:33.1 & UVES & 2 & 2024-06-08 & 238 \\ \hline
HD\,137927 & 21:03:54.25 & -11:27:48.5 & UVES & 2 & 2024-05-23 & 265 \\ \hline
HD\,138871 & 22:54:47.15 & +24:26:32.9 & UVES & 2 & 2024-06-22 & 239 \\ \hline
HD\,141705 & 21:54:48.95 & +01:13:48.9 & UVES & 2 & 2024-06-22 & 446 \\ \hline
HD\,142220 & 21:53:37.45 & +00:08:17.3 & UVES & 2 & 2024-06-22 & 264 \\ \hline
HD\,142705 & 23:16:9.95 & +22:39:43.6 & UVES & 2 & 2024-04-03 & 348 \\ \hline
HD\,150990 & 22:56:17.69 & +03:38:41.5 & UVES & 4 & 2024-07-31 to 2024-08-02 & 306 \\ \hline
HD\,152192 & 23:28:57.20 & +08:30:12.5 & UVES & 2 & 2024-07-31 & 360 \\ \hline
HD\,156248 & 00:47:13.16 & +15:18:6.8 & UVES & 2 & 2024-08-17 & 429 \\ \hline
HD\,15657 & 11:53:52.89 & -59:39:31.9 & UVES & 2 & 2024-06-27 & 149 \\ \hline
HD\,156622 & 23:00:46.30 & -03:36:36.2 & UVES & 2 & 2024-06-09 & 206 \\ \hline
HD\,163799 & 00:28:39.52 & +00:45:32.8 & UVES & 2 & 2024-08-03 & 314 \\ \hline
HD\,167307 & 02:08:22.55 & +09:56:11.3 & UVES & 2 & 2024-08-11 & 247 \\ \hline
HD\,173891 & 23:57:4.40 & -15:24:7.2 & UVES & 2 & 2024-06-28 & 277 \\ \hline
HD\,176425 & 23:40:8.38 & -19:36:11.9 & UVES & 2 & 2024-05-06 & 278 \\ \hline
HD\,178953 & 02:22:55.10 & -03:56:10.2 & UVES & 2 & 2024-06-28 & 246 \\ \hline
HD\,182621 & 02:39:34.24 & -05:59:19.04 & UVES & 2 & 2024-06-28 & 200 \\ \hline
HD\,198534 & 22:26:28.46 & -38:28:6.5 & UVES & 2 & 2024-06-27 & 158 \\ \hline
HD\,221272 & 05:55:19.45 & -52:11:35.1 & UVES & 2 & 2024-08-01 & 295 \\ \hline
HD\,230787 & 03:13:57.66 & +02:58:56.4 & UVES & 2 & 2024-08-12 & 286 \\ \hline
HD\,29929 & 16:45:35.95 & -41:25:50.3 & UVES & 2 & 2024-07-24 & 329 \\ \hline
HD\,307428 & 18:54:43.42 & -05:41:15.7 & UVES & 2 & 2024-05-22 & 227 \\ \hline
HD\,345199 & 04:01:37.43 & -02:12:27.4 & UVES & 2 & 2024-05-20 & 297 \\ \hline
HD\,35714 & 13:05:45.69 & -15:08:9.1 & UVES & 2 & 2024-08-12 & 310 \\ \hline
HD\,70007 & 17:33:20.49 & -06:23:16.5& UVES & 2 & 2024-05-23 & 75 \\ \hline
HD\,76516 & 17:38:44.44 & +00:47:52.8 & UVES & 2 & 2024-05-22 & 260 \\ \hline
HD\,80951 & 19:20:56.60 & -17:35:41.5 & UVES & 2 & 2024-05-21 & 178 \\ \hline
HD\,87122 & 18:43:3.71 & +00:43:59.9 & UVES & 2 & 2024-05-18 & 262 \\ \hline
HD\,92316 & 19:28:21.60 & -10:26:16.6 & UVES & 2 & 2024-05-21 & 207 \\ \hline
V*\,IN\,Lup & 22:33:35.06 & +10:52:3.5 & UVES & 2 & 2024-07-31 & 476 \\ \hline
V*\,KN\,Vel & 18:21:25.53 & +02:49:49.4 & UVES & 2 & 2024-05-21 & 189 \\ \hline
\end{longtable}

\end{onecolumn}

\section{Photosphere fits}\label{Appendix:B}

\begin{figure}
\centering
\includegraphics[width=0.9\columnwidth]{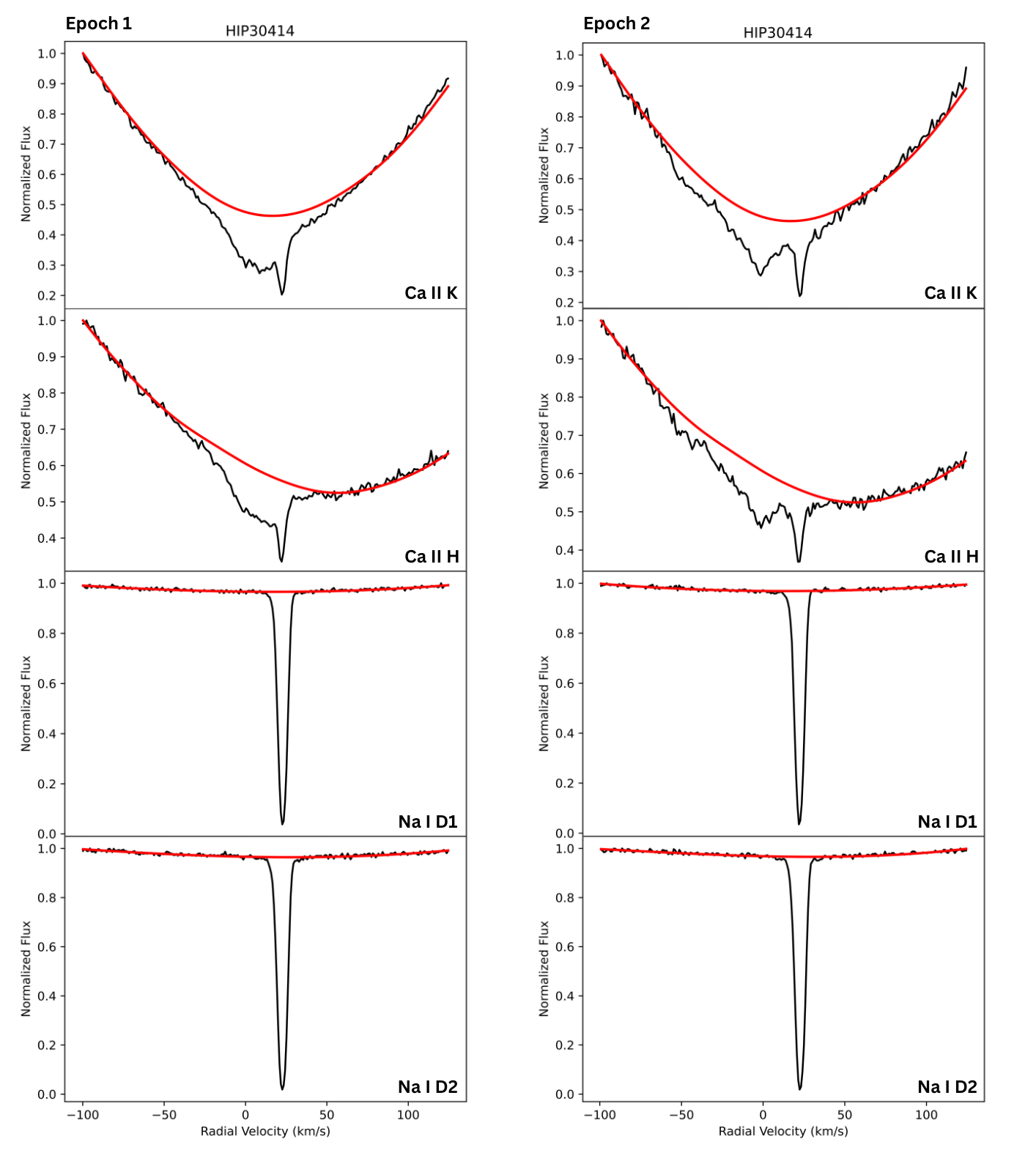}
\caption{\label{fig:HIP30414_appendix}Photospheric spectrum of HIP\,30414 in Ca\,\textsc{ii} K \& H and Na\,\textsc{i} D1 \& D2 lines. Columns are two different epochs we observed, showing variability between them. Black line is the flux, while red curve is the fitted Kurucz photosphere model used to isolate narrow features. The model parameters for Ca\,\textsc{ii} K are: temperature 7440\,K, log(g) 4.2, [Fe/H] 0, vsin(i) 130\,km\,s$^{-1}$; for Ca\,\textsc{ii} H: temperature 7400\,K, log(g) 4.2, [Fe/H] 0, vsin(i) 135\,km\,s$^{-1}$.}
\end{figure}

\begin{figure} 
\centering
\includegraphics[width=0.9\columnwidth]{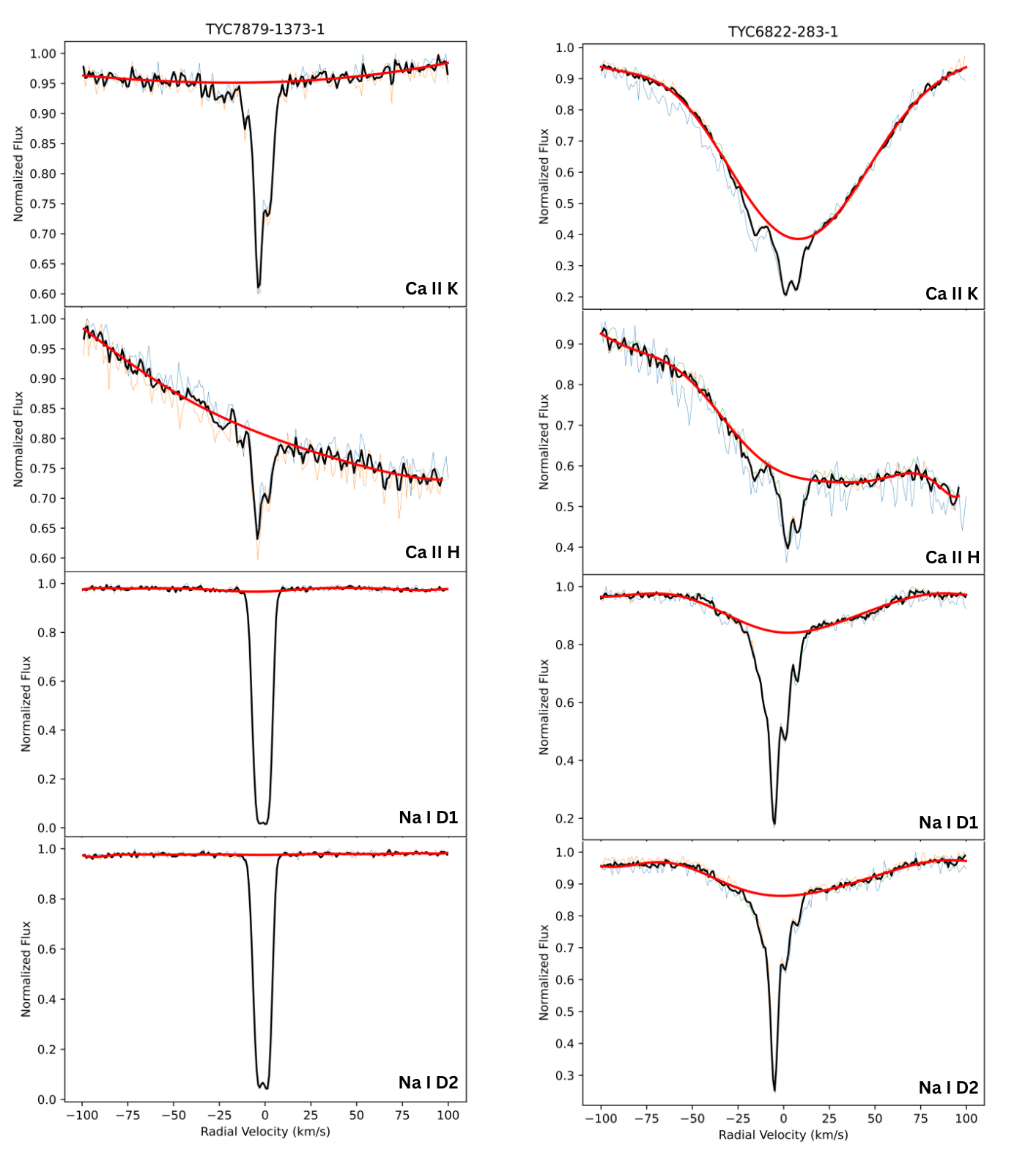}
\caption{\label{fig:TYC7879-1373-1_TYC6822-283-1_appendix}\textbf{Left: }Photospheric spectrum of TYC\,7879-1373-1, in Ca\,\textsc{ii} K \& H and Na\,\textsc{i} D1 \& D2 lines, as normalized flux vs radial velocity [km\,s$^{-1}$]. Black line is the median flux of all epochs, while red curve is the photospheric absorption fitted with a polynomial used to isolate narrow features. \textbf{Right: }Same but for TYC\,6822-283-1.}
\end{figure}


\bsp	
\label{lastpage}
\end{document}